\documentclass[fleqn,usenatbib]{mnras}

\usepackage{newtxtext,newtxmath}
\usepackage{graphicx}
\usepackage{mathtools}
\usepackage{amsmath}
\usepackage[T1]{fontenc}
\usepackage{physics}
\usepackage{bbold}
\usepackage{geometry}
\usepackage{scalerel}
\usepackage[scientific-notation=true]{siunitx} %for \num

\usepackage[titlenumbered,linesnumbered,ruled]{algorithm2e}
\usepackage{xcolor}
\SetCommentSty{mycommfont}

\newcommand{\unit}{\mathbb{1}}
\newcommand{\rstar}{\vb*{r}^{\ast}}
\newcommand{\sstar}{\vb*{s}^{\ast}}
\newcommand{\xstar}{\vb*{x}^{\ast}}

\newcommand{\rvline}{\hspace*{-\arraycolsep}\vline\hspace*{-\arraycolsep}}

\DeclareMathOperator*{\argmax}{argmax}

\newcommand{\lnb}[1]{%
  \ln\left[#1\right]%
}

\newcommand{\carpool}{CWAV20}
\newcommand{\carpcov}{CW21}

\newcommand{\indep}{\rotatebox[origin=c]{90}{$\models$}}

\title{Bayesian Control Variates for optimal covariance estimation with pairs of simulations and surrogates}

\author[Nicolas Chartier and Benjamin D.~Wandelt]{
\parbox{\textwidth}{
\LARGE
Nicolas Chartier$^{1, 2}$
and Benjamin D.\ Wandelt $^{2,3}$
}
\vspace{0.4cm}
\\
$^{1}$Laboratoire de Physique de l'\'{E}cole Normale Sup\'{e}rieure, ENS, Universite PSL, CNRS, Sorbonne Universit\'{e}, Universit\'{e} Paris Cit\'{e}, F-75005 Paris, France\\
$^{2}$Sorbonne Universit\'{e}, CNRS, UMR 7095, Institut d'Astrophysique de Paris, 98 bis bd Arago, 75014 Paris, France\\
$^{3}$Center for Computational Astrophysics, Flatiron Institute, 162 5th Avenue, New York, NY 10010, USA
}

\date{Accepted XXX. Received YYY; in original form ZZZ}
\pubyear{2022}

\begin{document}

\maketitle

\begin{abstract}
Predictions of the mean and covariance matrix of summary statistics are critical for confronting cosmological theories with observations, not least for likelihood approximations and parameter inference. The price to pay for accurate estimates is the extreme cost of running $N$-body and hydrodynamics simulations. Approximate solvers, or \textit{surrogates}, greatly reduce the computational cost but can introduce significant biases, for example in the non-linear regime of cosmic structure growth. We propose "CARPool Bayes", an approach to solve the inference problem for both the means and covariances using a combination of simulations and surrogates. Our framework allows incorporating prior information for the mean and covariance. We derive closed-form solutions for \textit{Maximum A Posteriori} covariance estimates that are efficient Bayesian shrinkage estimators, guarantee positive semi-definiteness, and can optionally leverage analytical covariance approximations. We discuss choices of the prior and propose a simple procedure for obtaining optimal prior hyperparameter values with a small set of test simulations. We test our method by estimating the covariances of clustering statistics of \texttt{GADGET-III} $N$-body simulations at redshift $z=0.5$ using surrogates from a 100-1000$\times$ faster particle-mesh code. Taking the sample covariance from 15,000 simulations as the truth, and using an empirical Bayes prior with diagonal blocks, our estimator produces nearly identical Fisher matrix contours for $\Lambda$CDM parameters  using only $15$ simulations of the non-linear dark matter power spectrum. In this case the number of simulations is so small that the sample covariance would be degenerate. We show cases where even with a na\"ive prior our method still improves the estimate. Our framework is applicable to a wide range of cosmological and astrophysical problems where fast surrogates are available.
\end{abstract}


% \bdw{Things to add: \begin{itemize}
%     \item discussion of limit of exactly known surrogate statistics (eqiuvalent to infinite number of surrogate samples) (19/01/22: result is there now, in Appendix A. needs discussion in main body);
%     \item sampling \nc{Discussion in conclusion}
%     \item  uncertainty estimate for the covariance: e.g. hessian/ Fisher matrix.
%     \item more example results \nc{DONE (Bk in appendix C)}
%     \item include factor to make precision estimates unbiased (plots done, update in paper) \nc{in appendix}
%     \item do we have a correction factor a la Kaufman/Hartlap when we compute precision from our estimate.- (19/01/22: guess based on form of estimator: "hartlap" with $n\rightarrow n_s+n_r+n_p$) \nc{DONE, but needs to be clearer}
%     \item mention that this is a unification for both means and covariances, but we focus on the covariance estimation here because we see the largest gains. Limit $n_p\rightarrow  \infty$ for empirical Bayes with block prior with diagonal blocks recovers the case in the first CARPool paper.
%     \item \nc{cite the DESI paper using CARPool}
% \end{itemize} }

\section{Introduction}\label{sec:intro}
To study the large-scale structure of the universe and cosmic growth history in the era of data-driven cosmology, one needs to accurately model the statistical properties of observables in order to infer cosmological parameters constraints from surveys. The covariance matrix $\vb*{\Sigma}$ of a summary statistics vector, such as the matter power spectrum across different wave-numbers, and most importantly its inverse ---the \textit{precision} matrix---are paramount to extracting low-dimensional summaries, building inference frameworks or computing likelihood approximations from mock catalogues (\cite{moped,refId0,2009ApJ...700..479T,2012MNRAS.426.1262H,2013PhRvD..88f3537D,2013MNRAS.431.3349H,2014MNRAS.442.2728T,2014MNRAS.439.2531P,10.1093/mnras/stu2190,2014IAUS..306...99J,alsingwandelt2018,2019A&A...631A.160H,2020PhRvD.102h3514H,2020PhRvD.102l3521W,2021A&A...653A..19G}).

The most trusted yet costly method to compute the covariance matrix of large-scale structure clustering statistics consists in generating mock realizations of survey observables with intensive $N$-body simulations -- or even hydrodynamical simulations for certain applications -- that mimic the conditions of observational data sampling in terms of redshift, sky area, volume, etc,... We then use the samples to compute the unbiased and positive definite sample covariance estimator, but getting a reliable estimate requires many realizations, especially if we need the precision matrix for the estimation of parameter confidence bounds.

To reduce the computational cost of generating simulation samples various parallel, distributed-memory $N$-body solvers have been developed sometimes with GPU-acceleration (
\citet{2005MNRAS.364.1105S} (GADGET), \citet{2009PASJ...61.1319I} (GreeM), \citet{2013arXiv1310.4502W} (2HOT), \citet{2013MNRAS.436..540H} ( CUBEP$^{3}$M), \citet{phdAbacus} (Abacus), \citet{2016NewA...42...49H} (HACC), \citet{2017ComAC...4....2P} (PKDGRAV3), \citet{2018ApJS..237...24Y} and \citet{2020arXiv200303931C} (CUBE)). 
Relying solely on massively parallel computing  to tackle next-generation observational datasets appears impractical given our time, memory, and energy resources since thousands of simulations are needed to produce sufficiently accurate cosmological parameter constraints (see for instance \citet{2016MNRAS.458.4462B}).

For this reason, cosmologists have been searching for  alternatives to running a large number of $N$-body simulations for a particular cosmological model.

On the theoretical side, analytical computations give covariance matrices that have little or no Monte Carlo noise but approximate and only valid for some assumptions on the data model. Such computations typically exploit the Gaussian limit and/or deviations from Gaussianity of the covariance \citep{2019MNRAS.490.5931P,2019JCAP...01..016L,2020MNRAS.491.3290P} or stem from perturbation theory \citep{Mohammed:2014lja,2017MNRAS.466..780M}. For reviews of methods motivated by theoretical predictions, refer to \citet{Bernardeau:2001qr} and \citet{Desjacques:2016bnm}.

On the computational side, researchers have developed approximate solvers which are much faster than full $N$-body or hydrodynamical codes at the cost of coarser computations and simplifications that reduce the overall accuracy with respect to intensive solutions, especially at small scales. An important part of these approximate solvers use Lagrangian Perturbation Theory (LPT) within a low-fidelity Particle-Mesh (PM) framework: \citet{Tassev_2013} (COLA), \citet{2015arXiv150207751T} (sCOLA) implemented by \citet{2020A&A...639A..91L}, \citet{2016MNRAS.463.2273F} (FastPM) available in a distributed version by \citet{2020arXiv201011847M}, \citet{2014MNRAS.437.2594W} (QPM), and \citet{2014MNRAS.439L..21K} (PATCHY), to name a few. Methods based on low order LPT predictions provide numerous fast structure formation statistics for cosmology:  \citet{2002MNRAS.329..629S} (PTHalos), \citet{2012JCAP...04..013T} and \citet{2013MNRAS.433.2389M} building upon the work of \citet{pino} (PINOCCHIO), or \citet{2015MNRAS.446.2621C} (EZmocks).

An increasingly popular approach, based upon optimization, is to construct mathematical models--\textit{emulators}-- that directly predict summary statistics for specific cosmologies and parameters and of which the free-parameters were previously determined through training with a specific loss function and, most importantly, simulation suites covering an appropriate range of the space of the upcoming input data \citep{2019arXiv190713167M, Zhai_2019,  McClintock_2019, DeRose_2019,2019MNRAS.490..331L, 2020arXiv200108055K, 2020arXiv200406245A, 2020ApJS..249....5A, 2021PhRvD.103d3526R,2021JCAP...05..033P}.
A large proportion of the underlying mathematical models of emulators are trained Neural Networks architectures \citep{2020arXiv201110577L,2020arXiv201108271R,2020arXiv201200240A, 2021arXiv210909747V, cosmopower2022} that produce summary statistics, and some have been specifically designed to output matter density fields from input initial conditions, or even snapshots of low-resolution $N$-body simulations with particles positions and velocities  \citep{He_2019, 2020arXiv201002926D,Kodi_Ramanah_2020}. Recently, \citet{2021arXiv210412864M} proposed a solution to the inverse problem of estimating the initial density field of the Early Universe : they combine a differentiable $N$-body solver with a Recurrent Neural Network architecture (RNN) to build a tractable inference scheme. Also, \citet{2021arXiv211002983H} (HIFLOW) trained an emulator and are able to produce 2D neutral Hydrogen maps conditioned on cosmology.

As a consequence of the growing enthusiasm for \textit{Machine Learning} solvers we have seen the production of massive simulation suites -- \cite{2018ApJS..236...43G}, \citet{2020ApJS..250....2V}, \citet{2021ApJ...915...71V} and \citet{2021arXiv210910915V} -- that more and more often aim specifically at providing ways to train various emulators and models. Any trained model suffers from two main drawbacks: namely the need for many training simulations and the subsequent limitation of the model to generalize by the parameter range of the training set; and the absence of guarantee for unbiasedness of the predictions with respect to full $N$-body or hydrodynamical outputs.

All the fast solvers described above -- which we will refer to collectively as \textit{surrogates} --  trade the accuracy of full $N$-body mocks, especially in the non-linear regime at small scales, for computational speed and memory gains. As a consequence, parameters constraints derived from surrogates only do not match the reliability and accuracy needed for upcoming surveys. For experiments, see the studies by \citet{2019MNRAS.482.1786L}, \citet{2019MNRAS.485.2806B} and \citet{2019MNRAS.482.4883C},  where statistical biases in parameters estimation using covariance matrices from surrogates range up to $10-20\%$ higher than with covariances computed from full $N$-body solvers.

Another approach is to attempt to reduce the number of needed simulations by modifying the statistical estimator of the covariance matrix. Numerous studies have been encouraging the use of new methods in order to deal with future surveys large data sets: \textit{covariance tapering} in \citet{10.1093/mnras/stv2259} who demonstrated the ability to reduce the confidence intervals of parameters without adding bias, fitting a theoretical model with mock samples \citep{2016MNRAS.457..993P}, jackknife resampling for the covariance \citep{2016arXiv160600233E, 2020arXiv200413436F}, reducing the number of simulations by using both theoretical and simulated covariances \citep{2019MNRAS.483..189H}, combining an empirical covariance with a simple target via (non-)linear shrinkage \citep{2008MNRAS.389..766P,2017MNRAS.466L..83J}.
As hinted at above, precision matrix estimation is the elephant in the room when it comes to undesirable effects  -- parameters shifts... -- of poor conditioning onto parameter constraints. Among the recent papers that deal with these limits and means to overcome (some of) them, the reader can refer to \citet{2013MNRAS.432.1928T} who show how the accuracy of the precision matrix impacts parameters constraints in the case of Gaussian-distributed weak lensing power spectra, the \textit{precision matrix expansion} method from \citet{2018MNRAS.473.4150F}, \citet{2018MNRAS.473.2355S} who show the limit of a Gaussian likelihood to derive parameter constraints, the Appendix B of \citet{2021PhRvD.103d3508P} that details parameter shifts stemming from a noisy covariance estimate, \citet{2021arXiv210810402P} who choose a specific covariance prior in a Bayesian framework, and also the Dark Energy Survey (DES) Year 3 results from \citet{2021MNRAS.tmp.2208F}.

Variance reduction methods allow to exploit the accuracy of $N$-body solvers while dramatically lowering the number of required samples to compute robust moments estimators. \citet{2021MNRAS.500..259S}, for example, combined different lines of sight in redshift space and lowered the variance of the quadrupole estimator of the two-point clustering statistic by more than one third.

\citet{2016PhRvD..93j3519P}, \citet{2016MNRAS.462L...1A}, and \citet{Villaescusa_Navarro_2018} discuss variance reduction with simulation pairs having special initial conditions. The technique allows to estimate the mean of statistics such as the power spectrum, the monopole and quadrupole of the redshift-space correlation functions or the halo mass function faster by a factor of more than $50$. The induced bias, however, on certain higher-order $N$-point functions renders the method not adapted to covariance estimation.

In \citet{carpool} and \citet{carpcov} (\carpool\ and \carpcov\ from now on), we developed the Convergence Acceleration by Regression and Pooling (CARPool) method, a general approach to reducing the number of simulations needed for low variance and explicitly unbiased estimates of clustering statistics moments. \carpool\ demonstrated a dramatic reduction of the number of simulations required to estimate the mean of a given statistic by exploiting the variance reduction principle known as \textit{control variates}. The key idea is to combine a small number of costly simulations with a large number of correlated \textit{surrogates}. Very recently, \citet{desiCARPool} tested the CARPool principle to estimate the mean of the two-point and three-point clustering statistics of halos, in order to prepare the high-resolution simulations needed for the Dark Energy spectroscopic Instrument (DESI). By pairing \textit{AbacusSummit} suite \citep{abacusSummit} simulations with FastPM approximations, they found $ \approx 100$ times smaller variances with CARPool at scales $k \leq 0.3$ $h {\rm Mpc^{-1}}$ than with high-resolution simulations alone. Additionally, the extension of the method to different cosmologies (one or very few simulations of the cosmologies of interest paired with a "primary cosmology" as the surrogate) resulted in an increase of the \textit{effective volume} by $\approx 20$ times.
In \carpcov, we extended the principle to covariance estimation by applying the variance reduction approach to individual elements of a symmetric matrix, and we assessed the covariance estimates by deriving cosmological parameters confidence intervals with the Fisher matrix (using the precision). With this straightforward approach we found significant improvement in many cases, but a definite drawback was that positive-definiteness of the covariance estimate is not guaranteed. The main reason for this was because the covariance matrix was treated as a first order statistic for the methods in \carpool\ to be directly applicable.

In this paper, to circumvent this drawback, we frame the problem as a Bayesian inference of simulation means and covariances  when a (typically small) set of pairs of simulations and surrogates are available in addition to a (typically large) set of unpaired fast surrogates. We derive closed-form Maximum A Posteriori (MAP) estimators of the covariance of the simulation statistics that incorporate the information brought by the surrogates and the prior, and test the estimates by comparing the resulting confidence bounds for a $\Lambda$CDM cosmology with the true bounds. %We also provide an exact algorithm to produce independent posterior samples of the covariance matrix. 
The results in this paper are very general and can apply to any summary statistics from simulations. For this reason, we motivate the study with an introductory example in section \ref{sec:illustration} before explaining the notations and derivations in section \ref{sec:theory}. We show several example applications to large scale structure statistics in section \ref{sec:exp} and we conclude and discuss the implications of our work in section \ref{sec:conclusion}.

\section{Illustrative example}\label{sec:illustration}
% Among the stakes of a forecast analysis, there is the ability to predict the constraining power on cosmology from an observable we will measure in the context of a given survey (galaxy survey, footprint on the sky, etc...).
% In other words: "If I spend such time and resources to measure my observable, how much will I be able to learn about physics, about the Universe?".
% For $\Lambda$CDM cosmology, the vector of parameters is $\boldsymbol{\theta} = \left(\Omega_m, \Omega_b,h,n_s,\sigma_8,M_{\nu}\right)^{\boldsymbol{T}}$, and $\mathbb{E}_{\boldsymbol{\theta}}\left[\boldsymbol{y}\right] = \boldsymbol{\mu}(\boldsymbol{\theta})$ is the expectation of $\boldsymbol{y}$ for fixed parameters $\boldsymbol{\theta}$.
% The partial derivatives of the statistics are estimated numerically using finite differences from $500$  \textit{Quijote} simulations for each varying parameter exactly as in \citet{2020ApJS..250....2V}.

\begin{figure*}
    \includegraphics[width=\textwidth]{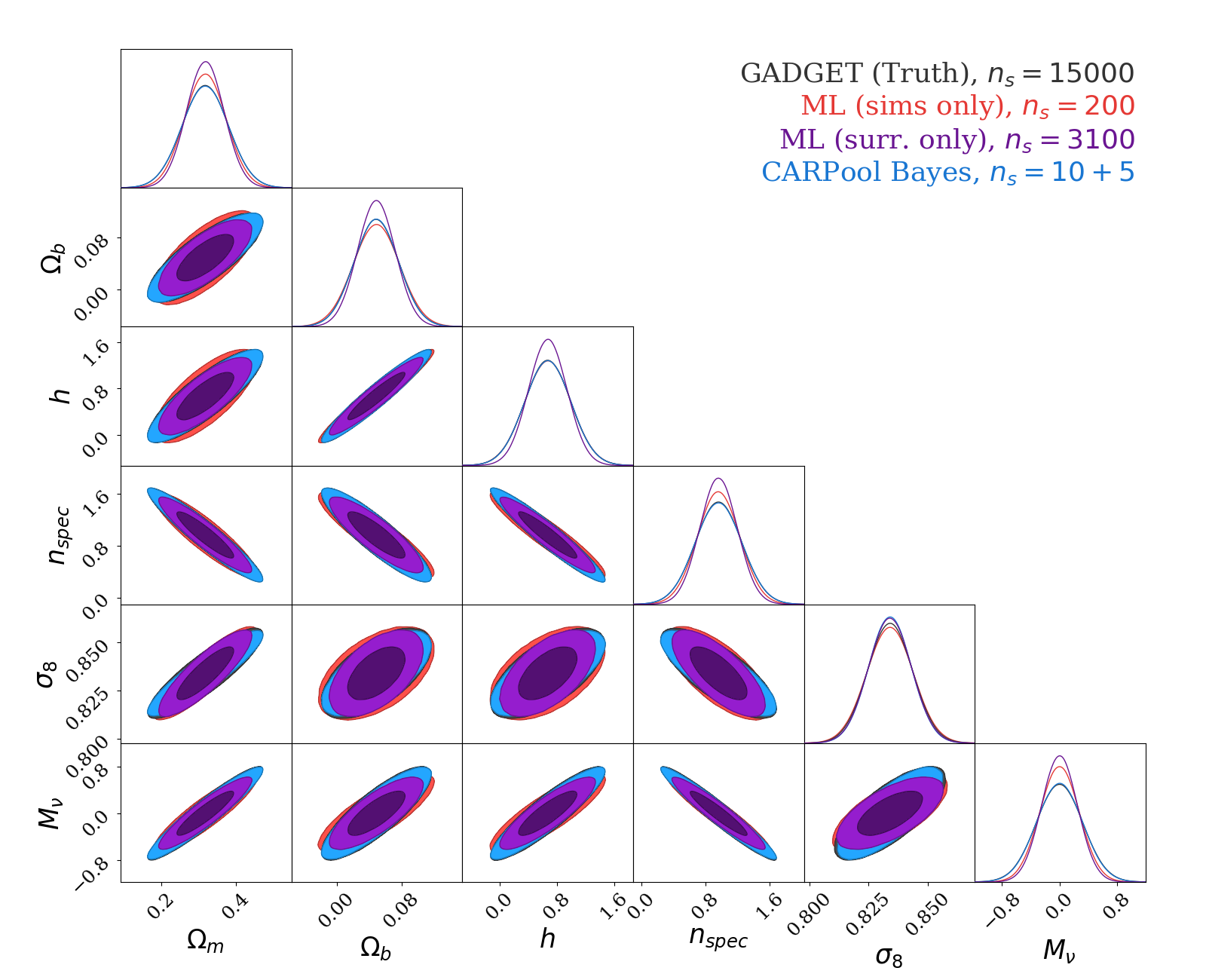}
    \caption{Illustrating the power of Bayesian control variates using the confidence contours of the cosmological parameters computed using the Fisher matrix based on the estimated matter power spectrum covariance matrix. The "truth" designates the confidence regions (black) from the sample covariance matrix of 15,000 $N$-body simulations, and the parameter means are set to the $\Lambda$CDM model used in the simulations. The contours are overlapped nearly perfectly by the light blue when the covariance in the Fisher matrix is computed using only $15$ simulations with our CARPool Bayes MAP estimator ($10$ simulations and $5$ for setting a prior hyperparameter, see section \ref{sec:priorHyper}). The sample covariance (ML) estimator based on many more simulations than ours gives less accurate contours. Contours based on $3100$ COLA surrogates alone are rotated and too small showing that the surrogates alone are inaccurate. Detailed discussion in the text and in section \ref{sec:exp}.}
    \label{fig:fisherPk}
\end{figure*}

Imagine having a simulation code to compute the evolution of collisionless dark matter particles in an expanding $\Lambda$CDM universe, within a simulation volume mimicking the observational conditions of some future survey. We would like to ask: "What amount of information the clustering statistics of the large-scale structure carry about the cosmological parameters? By which amount will we be able to constrain cosmological parameters with said statistics?" Let's say we try with the two-point correlation function in Fourier space, \textit{i.e.,} the (dark matter) power spectrum. For each of $n_s$ runs, with different random seeds for the initial conditions, labelled $i, 1 \leq i \leq n_s$, the output is the vector $\vb*{s}_i$ of $p_s = 158$ power spectrum bins up to $k_\mathrm{max} \approx 1.0$ $h {\rm Mpc^{-1}}$. {We will introduce the detailed notation in section \ref{sec:notations}.}

Under the hypothesis that the observable is sampled from a Multivariate Normal (MVN) distribution and that the covariance matrix does not depend on the parameters, the Fisher matrix for $d$ parameters is the symmetric matrix of size $(d,d)$ 
\begin{equation}
    \begin{aligned}
    \mathcal{F}_{ij} = \left(\frac{\partial \boldsymbol{\mu(\boldsymbol{\theta})}}{\partial \theta_i}\right)^{\boldsymbol{T}}\boldsymbol{\Sigma_{yy}^{-1}}\left(\frac{\partial \boldsymbol{\mu(\boldsymbol{\theta})}}{\partial \theta_j}\right)\,;
    \end{aligned}\label{eq:fisherMVG}
\end{equation}
hence the importance of having an accurate estimate of the covariance matrix and its inverse,  the precision matrix.
Then, for a parameter $\theta_i$, the Cramér-Rao inequality gives the lower-bound, marginalized over the remaining parameters, for the variance of an unbiased estimator of $\theta_i$:
\begin{equation}\label{eq:cramerRao}
    \sigma^2_{\theta_i} \geq {\left[\mathcal{F}^{-1}\right]_{ii}}\,.
\end{equation}

To get an accurate estimate of the confidence bounds for the parameters requires an accurate estimate  of the covariance matrix $\vb*{\Sigma_{ss}}$. Using the standard sample covariance estimator (or  maximum likelihood estimator)  we would expect to need thousands of simulations costing $\mathcal{O}(10^7)$ CPU hours, much like in the \textit{Quijote} suite \citep{2020ApJS..250....2V}.  

But we have at our disposal a much faster surrogate solver that uses approximations from a Lagrangian fluid description of the dark matter field to produce fast but unfortunately biased approximations of this power spectrum. In this paper we show how to leverage these fast surrogates to obtain accurate estimates of the means and covariance of the summary statistics while reducing the required number of simulations by orders of magnitude. 

Figure \ref{fig:fisherPk} illustrates the take-home message of this work. It shows the predicted marginal confidence regions of $\Lambda$CDM cosmological parameters \footnote{We use $n_\textit{spec}$ as the spectral index not to induce confusion with the number of simulations $n_s$ used in the paper.}  computed using  different estimates of the power spectrum covariance.  The case labeled "Truth" uses the standard Maximum Likelihood Estimate (MLE) of the covariance matrix from 15,000 full simulations. This "Truth" case is hardly visible because the contours are nearly perfectly overlapped by the "CARPool Bayes" case that uses only $15$ simulations in combination with fast surrogates (noted as $10+5$ simulations, the second-term being the number of test simulations used to set a prior hyperparameter; see discussion in section \ref{sec:priorHyper}). This is one of the Bayesian covariance estimators we develop in this paper.
These two cases are to be compared with the  "ML (sims only)" case showing the standard MLE of the covariance matrix from $200$ simulations but without surrogates. The case labeled  "ML (surr. only)" illustrates that relying on the surrogates alone results in biased estimates of the size and orientation of the contours.
\footnote{We correct the bias of the precision matrix computed by inverting the standard sample covariance matrix estimator in equation \eqref{eq:fisherMVG} with the so-called "Hartlap factor" (see section \ref{sec:Hartlap} for a reminder) when using sample covariances, i.e. for "ML (sims only)", "ML (surr. only)" and also for the truth even if the correction is small. We do not use any correction when using the "CARPool Bayes" estimate, a point which we discuss in section \ref{sec:ourHartlap}.}

Figure \ref{fig:fisherPk} emphasizes the potential of the Bayesian formulation of the CARPool approach that we develop in detail in the following. Readers mostly interested in applications and numerical examples can skip to section 
\ref{sec:exp}.

\section{Bayesian Inference of Covariance  from Simulation-Surrogate Pairs}\label{sec:theory}
We wish to estimate the covariance matrix of the summary statistics $\vb*{s}$, $\mathrm{dim}(\vb*{s})=p_s$ from accurate, expensive simulations. We also have access to a fast surrogate solver, $\vb*{r}$, $\mathrm{dim}(\vb*{r})=p_r$, which we would not rely on alone. Inspired by CARPool, we build estimators to exploit both simulation and surrogate statistics, with the main goal of reducing the number of intensive simulations we have to run.

\subsection{Definitions and notations}\label{sec:notations}

With simulation summary statistics samples $\vb*{s}_i$, $i=1,\dots,n_s$ the standard approach to estimating the covariance matrix $\vb*{\Sigma_{ss}}$ is to compute
\begin{align}\label{eq:sampCov}
    \vb*{\widehat{\Sigma}_{ss}} &= \frac{\gamma}{n_s} \sum_{i=1}^{n_s} \left( \vb*{s}_i -\vb*{\bar{s}} \right) \left( \vb*{s}_i -\vb*{\bar{s}} \right)^{\vb*{T}}\\
    \vb*{\bar{s}} &= \frac{1}{n_s}\sum_{i=1}^{n_s}\vb*{s}_i\,,\nonumber
\end{align}
the \textit{Maximum-Likelihood} (ML) estimator given a Multivariate Normal (MVN) likelihood function when $\gamma=1$. To get an unbiased estimator, we use Bessel's correction factor $\gamma= n_s/(n_s -1)$ in the ML estimator for the covariance. Equation \eqref{eq:sampCov} needs many samples to provide a high-quality estimate: as a matter of fact, the convergence of the smallest eigenvalues is slow \citep{BaiYin} and these eigenvalues will dominate the precision matrix and impact parameter parameter constraints (\citet{2013MNRAS.432.1928T}, \citet{2016MNRAS.458.4462B}).

Now we add surrogates. The goal is to build a Bayesian model for the covariance of the simulations but including whatever information is provided by the surrogates.
The set of surrogates $\vb*{r}_j$, $j=1,\dots,n_s + n_r$ comprises $n_s$ samples that are paired with the simulations, \textit{i.e.}, they were computed using the same random numbers, and $n_r$ additional unpaired surrogates.
We combine pairs of simulations and surrogates into a single  vector
\begin{align}
    \vb*{x} &\equiv\mqty(\vb*{s}, \vb*{r})^{\vb*{T}}
\end{align}
which implies a block matrix structure for the mean and covariance
\begin{align}
    \vb*{\mu} &\equiv \mathbb{E}\left[ \vb*{x}\right] =\mqty(\vb*{\mu_s}, \vb*{\mu_r})^{\vb*{T}}\nonumber\\
    \vb*{\Sigma}&\equiv\mqty(\vb*{\Sigma_{ss}}&\vb*{\Sigma_{sr}}\\\vb*{\Sigma_{rs}}&\vb*{\Sigma_{rr}})\nonumber.
\end{align}
Following the standard notation, we will denote the Schur complement as
\begin{equation}
    \left( \vb*{\Sigma}/\vb*{\Sigma}_{rr} \right)\equiv \vb*{\Sigma}_{ss}-\vb*{\Sigma}_{sr}\vb*{\Sigma}_{rr}^{-1}\vb*{\Sigma}_{rs}.
\end{equation}
$\mathcal{S}_p^{+}$ designates the space of symmetric positive-definite matrices, which is a subset of $\mathbb{R}^{p(p+1)/2}$.

For the $n_r$ \textit{unpaired} surrogates $\rstar$ we introduce the unobserved (and in fact non-existent) corresponding simulations $\sstar$ as latent variables and then treat them as \textit{missing data}. Again we combine  into a vector $\xstar \equiv (\sstar, \rstar)^{\vb*{T}}$ giving 
\begin{align}
    \vb*{s}_1, \dots, \vb*{s}_{n_s}  \qquad,\qquad &\sstar_1, \dots,  \sstar_{n_r} \nonumber \\
    \underbrace{\vb*{r}_1, \dots, \vb*{r}_{n_s}}_{\vb*{x}}  \qquad,\qquad &\underbrace{\rstar_{1}, \dots, \rstar_{n_r}}_{\xstar}. \nonumber
\end{align}

We will also distinguish the empirical counterparts of the surrogate moments according to whether they use all the $\vb*{r}$ available or just the paired ones, \textit{i.e.},
\begin{align}
    \vb*{\overline{r}}, \vb*{\widehat{\Sigma}_{rr}} &\longrightarrow \text{estimated from the unpaired set only;}\nonumber\\
 \vb*{\overline{r}^{\star}}, \vb*{\widehat{\Sigma}_{rr}^{\star}} &\longrightarrow \text{estimated  from both the paired and unpaired sets.} \nonumber
\end{align}
For instance,
\begin{equation}
    \vb*{\overline{r}^{\star}} = \frac{1}{n_s+n_r}\sum_{j=1}^{n_s+n_r}\vb*{r}_j\,,\nonumber
\end{equation}
where we do not differentiate the paired and unpaired surrogates for simplicity ($\vb*{r}_{j}=\rstar_{j-n_s}$ if $j \geq n_s+1$).

We recall the well-known result that the best prediction $\widehat{\sstar}$ for any $\sstar$ given $\rstar$ with no constraints (i.e we do not restrict the problem to the class of linear estimators) under the square loss of residuals coincides, when under a MVN distribution, with the linear regression:
\begin{align}
    &\mathcal{P}(\sstar|\rstar,\vb*{\Sigma})= MVN(\vb*{\mu}_{\sstar|\rstar}, \vb*{\Sigma}_{\sstar|\rstar}) \\
    &\widehat{\sstar} = \vb*{\mu}_{\sstar|\rstar}=\vb*{\Sigma}_{sr}\vb*{\Sigma}_{rr}^{-1}(\rstar - \vb*{\mu_r}) + \vb*{\mu_s}\nonumber \\
    &\vb*{\Sigma}_{\sstar|\rstar}=\left( \vb*{\Sigma}/\vb*{\Sigma}_{rr} \right)\nonumber
\end{align}

The regression matrix of $\vb*{s}$ given $\vb*{r}$ will appear from now on as
\begin{equation*}
    \vb*{B} \equiv \vb*{\Sigma}_{sr}\vb*{\Sigma}_{rr}^{-1}
\end{equation*}

For legibility, and without loss of generality, we will write all random vectors as zero-mean in the derivations such that for any sample $i$
\begin{equation*}
    \vb*{x}_i \leftarrow \vb*{x}_i - \vb*{\mu_x}.
\end{equation*}
The final equations serving as numerical recipes will include  the means explicitly.

With these notations, we now turn to inferring the simulation block of the covariance $\vb*{\Sigma_{ss}}$ with the help of surrogates, given (multiple realisations of) $\vb*{x}$ and $\xstar$. 

\subsection{Maximum-likelihood solution with surrogates}\label{sec:ML}
In a  Gaussian model, the  log-likelihood of $n_s$ \textit{independent and identically distributed} (iid) samples of  $\vb*{x}$ and $n_r$ iid samples $\xstar$ of simulation-surrogate pairs, is
\begin{align}
    &-2\lnb{\mathcal{L}(\{\vb*{x}\},\{\rstar\}| \vb*{\Sigma})}=(n_s+n_r)\lnb{ \det\left( \vb*{\Sigma}\right)}\label{eq:LogLikelihood}\\
    &+\sum_{i=1}^{n_{s}} \vb*{x}_{i}^{\vb*{T}} {\vb*{\Sigma}}^{-1}\vb*{x}_{i} +
     \sum_{i=1}^{n_{r}} \xstar_{i}{}^{\vb*{T}} {\vb*{\Sigma}}^{-1}\xstar_{i} + c_f\nonumber\,, 
\end{align}
where $c_f$ is the remaining constant of the likelihood for the full model including $\vb*{x}$ and $\xstar$.
Treating the simulations $\sstar$ in $\xstar$ as unobserved, latent variables we  use the Expectation Maximization (EM) approach \citep{10.2307/2984875}. While EM is typically an iterative algorithm that can be slow to converge, we show in Appendix~\ref{app:EM}  that we can find the Maximum Likelihood (ML) estimators of the mean and of the covariance from simulations and surrogates in  closed-form by computing the fixed point of the EM iterations. These are
\begin{align}
    \vb*{\widehat{B}} &= \vb*{\widehat{\Sigma}_{sr}}\vb*{\widehat{\Sigma}_{rr}}^{-1}\\
    \widehat{\vb*{\mu_{s|r}}}&=\overline{\vb*{s}} +  \vb*{\widehat{B}} \left(\overline{\vb*{r}}^{\vb*{\star}} - \overline{\vb*{r}}\right)\label{eq:musrML} \\
    \vb*{\widehat{\Sigma}_{ss}^{\scaleto{\mathrm{ML}}{3.5pt}}} &= \widehat{\left( \vb*{\Sigma}/\vb*{\Sigma}_{rr} \right)} + \vb*{\widehat{B}}\vb*{\widehat{\Sigma}_{rr}^{\star}}\vb*{\widehat{B}^{T}}\label{eq:covSSML}\\
    &=\vb*{\widehat{\Sigma}_{ss}}+ \vb*{\widehat{B}}\left(\vb*{\widehat{\Sigma}_{rr}^{\star}} -\vb*{\widehat{\Sigma}_{rr}}\right)\vb*{\widehat{B}^{T}}\nonumber\,,
\end{align}
where $\vb*{\widehat{\Sigma}_{ss}}$ is the sample covariance from equation \eqref{eq:sampCov} using simulations only. We provide a proof in Appendix~\ref{app:EM} that as long as the covariance of the surrogate is positive definite the ML estimate $\vb*{\widehat{\Sigma}_{ss}^{\scaleto{\mathrm{ML}}{3.5pt}}}$ is guaranteed to be positive (semi-)definite\footnote{\citet{Anderson1957} derived the same ML estimator by integrating out the $\sstar$ in Eq.~\eqref{eq:LogLikelihood} to obtain the marginal likelihood for the observed samples only
\begin{align}
    &-2\lnb{\mathcal{L}(\{\vb*{x}\},\{\rstar\}| \vb*{\Sigma})}=n_s\lnb{ \det\left( \vb*{\Sigma}\right)}+\sum_{i=1}^{n_{s}} \vb*{x}_{i}^{\vb*{T}} {\vb*{\Sigma}}^{-1}\vb*{x}_{i} \nonumber\\
  &+n_r\lnb{\det\left(\vb*{\Sigma_{rr}}\right)}+\sum_{j=1}^{n_{r}} \rstar_{j}{}^{\vb*{T}} \vb*{\Sigma_{rr}}^{-1}\rstar_{j} + c_m\nonumber\,,
\end{align}
with $c_m$ the remaining constant of the model with missing $\sstar$.}.

As we will show in section \ref{sec:exp}, this solution improves the estimated simulation covariance significantly with respect to the ML covariance computed from simulations only, Eq.~\eqref{eq:sampCov}. But the key ingredient for many applications is the precision matrix: computing optimal data combinations, least square estimators and optimal filtering. We will see that the dramatically underestimated smallest eigenvalues of the ML estimate of the covariance are critical. 

Fortunately, the Bayesian  approach allows us to  include priors amounting to a form of regularization, as we will show now.

\subsection{Inclusion of a Prior Information and Maximum A Posteriori (MAP) solutions}\label{sec:MAP}
A convenient prior to choose for the block covariance $\vb*{\Sigma}$, with $P \equiv p_s + p_r$, is the Inverse-Wishart ($\mathcal{W}^{-1}$) prior with hyperparameters $\vb*{\Psi} \in \mathcal{S}_P^{+}$, the scale matrix, and $\nu$, the number of degrees of freedom.
With $n_p \equiv \nu + P +1$ then
\begin{align}\label{eq:invWishart}
     &\mathcal{W}^{-1}(\vb*{\Sigma}|{\mathbf\Psi},\nu)={\frac {\det(\mathbf {\Psi })^{\nu /2}}{2^{\nu P/2}\Gamma _{P}({\frac {\nu }{2}})}}\det(\mathbf{\Sigma})^{-n_p/2}e^{-{\frac {1}{2}}\operatorname {tr} (\mathbf {\Psi } \vb*{\Sigma} ^{-1})}\\
     &\vb*{\Psi} \equiv\mqty(\vb*{\Psi_{ss}}&\vb*{\Psi_{sr}}\\\vb*{\Psi_{rs}}&\vb*{\Psi_{rr}})\nonumber\,,
\end{align}
where $\Gamma_P$ is the multivariate Gamma function.  $\mathcal{W}^{-1}(\vb*{\Sigma}|{\mathbf\Psi},\nu)$ has mode $\vb*{\Psi}/n_p$ for $n_p > 2P$. Its mean  $\vb*{\Psi}/(n_p - (2P + 2))$ exists if $n_p > 2P + 2$. In our problem, for any prior $\mathcal{P}(\vb*{\Sigma})$, the mode of the posterior distribution is located at the Maximum A Posteriori (MAP) estimate
\begin{equation}\label{eq:MAPproblem}
    \vb*{\widehat{\Sigma}_{ss}^{\scaleto{\mathrm{MAP}}{3.5pt}}} = \argmax_{\vb*{\Sigma_{ss}} \in \mathcal{S}_P^{+}}\left[ \mathcal{L}(\{\vb*{x}\},\{\rstar\}| \vb*{\Sigma}) \times \mathcal{P}(\vb*{\Sigma})\right]
\end{equation}
In order to get a $\vb*{\Sigma_{ss}}$ MAP estimate, we chose to study two approaches:
solving the MAP either for the whole $\vb*{\Sigma}$ matrix and $\mathcal{W}^{-1}$ prior (section \ref{sec:regML}), or for the "regression" parameters used to infer the $\vb*{\Sigma_{ss}}$ block, which amounts to dealing with the problem solved in \citet{Anderson1957} and reparametrizing the $\mathcal{W}^{-1}$ prior (section \ref{sec:newMAP}).

\subsubsection{MAP with prior on the block covariance $\vb*{\Sigma}$}\label{sec:regML}

We take $\mathcal{P}(\vb*{\Sigma}) = \mathcal{W}^{-1}(\vb*{\Sigma}|{\mathbf\Psi},\nu)$.
The derivation of the MAP estimator for the "full" covariance, in this case, bears similarity to the well-known proof that the Inverse-Wishart distribution is a conjugate prior for the covariance matrix under a MVN likelihood (where $\vb*{\Psi}$ becomes an additional factor of $\vb*{\Sigma}^{-1}$ in the trace factorization of the log-likelihood). 
In the absence of additional unpaired surrogates in equation \eqref{eq:LogLikelihood}, the MAP estimator for $\vb*{\Sigma}$ with the prior of equation \eqref{eq:invWishart} would match the classical result
\begin{equation}
    \vb*{\widehat{\Sigma}^{\Delta}} =\frac{n_s \vb*{\widehat{\Sigma}} + \vb*{\Psi}}{n_s + n_p} \equiv \mqty(\vb*{\widehat{\Sigma}_{ss}^{\Delta}}&\vb*{\widehat{\Sigma}_{sr}^{\Delta}}\\\vb*{\widehat{\Sigma}_{rs}^{\Delta}}&\vb*{\widehat{\Sigma}_{rr}^{\Delta}})\label{eq:SigmassDelta}
\end{equation}

The unpaired surrogate samples, in our case, can be used in addition to the standard $\vb*{\widehat{\Sigma}_{ss}^{\Delta}}$:

\begin{align}
    \vb*{\widehat{\Sigma}_{rr}^{\scaleto{\mathrm{MAP}}{3.5pt}}} &= \frac{(n_s + n_r) \vb*{\widehat{\Sigma}_{rr}^{\star}} + \vb*{\Psi_{rr}}}{n_s + n_r + n_p} = \frac{n_r \vb*{\widehat{\Sigma}_{rr}} + (n_s + n_p)\vb*{\widehat{\Sigma}_{rr}^{\Delta}}}{n_s + n_r + n_p}\label{eq:covRRMAP}\\
    \vb*{\widehat{B}_{\scaleto{\mathrm{MAP}}{3.5pt}}} &= \vb*{\widehat{\Sigma}_{sr}^{\Delta}} \vb*{\widehat{\Sigma}_{rr}^{\Delta}}^{-1}\\
    \vb*{\widehat{\mu_{s|r}^{\scaleto{\mathrm{MAP}}{3.5pt}}}}&= \overline{\vb*{s}} + \vb*{\widehat{B}_{\scaleto{\mathrm{MAP}}{3.5pt}}} \left(\overline{\vb*{r}}^{\vb*{\star}} - \overline{\vb*{r}}\right)\label{eq:muSRMAP}\\
    \vb*{\widehat{\Sigma}_{ss}^{\scaleto{\mathrm{MAP}}{3.5pt}}} &= \vb*{\widehat{\Sigma}_{s|r}^{\Delta}} + \vb*{\widehat{B}_{\scaleto{\mathrm{MAP}}{3.5pt}}}\vb*{\widehat{\Sigma}_{rr}^{\scaleto{\mathrm{MAP}}{3.5pt}}}\vb*{\widehat{B}_{\scaleto{\mathrm{MAP}}{3.5pt}}^{T}} \label{eq:covSSMAP}\\
    &= \vb*{\widehat{\Sigma}_{ss}^{\Delta}}+ \vb*{\widehat{B}_{\scaleto{\mathrm{MAP}}{3.5pt}}}\left(\vb*{\widehat{\Sigma}_{rr}^{\scaleto{\mathrm{MAP}}{3.5pt}}} - \vb*{\widehat{\Sigma}_{rr}^{\Delta}}\right)\vb*{\widehat{B}_{\scaleto{\mathrm{MAP}}{3.5pt}}^{T}}\nonumber\,,
\end{align}

Notice that in the absence of a prior, equation \eqref{eq:covSSMAP} reduces to equation \eqref{eq:covSSML} and to the standard result $\vb*{\widehat{\Sigma}_{ss}^{\Delta}}$ with no unpaired surrogates. 

Priors for the simulation and surrogate means could be trivially included as derived in Appendix A in \carpool.

Note that a simple limit of these equations exist for the case when the surrogate covariance is known exactly, \begin{equation}\label{eq:MAPlimit}
        %\widehat{\vb*{\Sigma}}^\text{MAP, $\vb*{\Sigma}_{rr}$}_{ss} =
        \vb*{\widehat{\Sigma}_{ss}^{\scaleto{\mathrm{MAP},\vb*{\Sigma}_{rr}}{5.0pt}}}=
        \vb*{\widehat{\Sigma}^{\Delta}_{ss}} + 
        \widehat{\vb{B}}_\text{MAP}
        \qty(\vb*{\Sigma}_{rr}-\widehat{\vb*{\Sigma}}_{rr}^{\Delta})
        \widehat{\vb{B}}_\text{MAP}^T.
\end{equation}
In Appendix \ref{sec:knownSurrCov}  we show that this result can be obtained by taking the limit of equation \eqref{eq:covSSMAP} for infinite number of surrogates.
In this case no unpaired surrogates need to be generated which can lead to significant savings when the computational expense for generating a large number of unpaired surrogates is not negligible compared to the simulation cost. In addition, any residual error in the estimate due to a limited number of surrogates is eliminated.

\subsubsection{MAP with prior on the regression parameters}\label{sec:newMAP}

A different approach is to solve the MAP for the parameters that allow to estimate $\vb*{\Sigma_{ss}} = \vb*{\Sigma_{s|r}} +  \vb*{B}\vb*{\Sigma_{rr}}\vb*{B^T}$, that is to say we use a prior for the joint distribution $\mathcal{P}(\vb*{B}, \vb*{\Sigma_{s|r}}, \vb*{\Sigma_{rr}})$ which is a reparametrization of the $p_s(p_s+1)/2 + p_r(p_r+1)/2 + p_sp_r$ parameters of $\mathcal{P}(\vb*{\Sigma})$. For that, we need the properties of the blocks of a covariance sampled from a $\mathcal{W}^{-1}(\vb*{\Sigma}|{\mathbf\Psi},\nu)$ distribution. A quick outline of the derivation appears in Appendix \ref{app:regressionMAP}. With $\vb*{B_{\Psi}} \equiv \vb*{\Psi_{sr}}\vb*{\Psi_{rr}}^{-1}$ we get

\begin{align}
    \vb*{\widehat{\Sigma}_{rr}^{\scaleto{\mathrm{MAP}}{3.5pt}}} &= \frac{(n_s + n_r) \vb*{\widehat{\Sigma}_{rr}^{\star}} + \vb*{\Psi_{rr}}}{n_s + n_r + \nu - p_s + p_r + 1}\label{eq:regRR}\\
    %\vb*{\widehat{\Sigma}_{s|r}} &= \sum_{i=1}^{n_s} \left(\vb*{s}_i - \vb*{\widehat{\mu_{s|r}^{\scaleto{\mathrm{MAP}}{3.5pt}}}}\right)\left(\vb*{s}_i - \vb*{\widehat{\mu_{s|r}^{\scaleto{\mathrm{MAP}}{3.5pt}}}}\right)^{\vb*{T}}\\
    \vb*{\widehat{B}_{\scaleto{\mathrm{MAP}}{3.5pt}}} &= \bigg[ \vb*{\Psi_{sr}} + \sum_{i=1}^{n_s}(\vb*{s}_i - \vb*{\mu_s})(\vb*{r}_i - \vb*{\mu_r})^{\vb*{T}} \bigg] \label{eq:regB}\\ &\times \bigg[\vb*{\Psi_{rr}} + \sum_{i=1}^{n_s}(\vb*{r}_i - \vb*{\mu_r})(\vb*{r}_i - \vb*{\mu_r})^{\vb*{T}} \bigg]^{-1}\nonumber\\
    \vb*{\widehat{\Sigma}_{s|r}^{\scaleto{\mathrm{MAP}}{3.5pt}}} &= \frac{n_s\vb*{\widehat{\Sigma}_{s|r}} + \vb*{\Psi_{s|r}} + \left(\vb*{\widehat{B}_{\scaleto{\mathrm{MAP}}{3.5pt}}} - \vb*{B_{\Psi}} \right)\vb*{\Psi_{rr}} \left( \vb*{\widehat{B}_{\scaleto{\mathrm{MAP}}{3.5pt}}} - \vb*{B_{\Psi}}\right)^{\vb*{T}}}{\nu + n_s + 2ps + 1} \label{eq:regSR}\\
    \vb*{\widehat{\Sigma}_{ss}^{\scaleto{\mathrm{MAP}}{3.5pt}}} &= \vb*{\widehat{\Sigma}_{s|r}^{\scaleto{\mathrm{MAP}}{3.5pt}}} + \vb*{\widehat{B}_{\scaleto{\mathrm{MAP}}{3.5pt}}}\vb*{\widehat{\Sigma}_{rr}^{\scaleto{\mathrm{MAP}}{3.5pt}}}\vb*{\widehat{B}_{\scaleto{\mathrm{MAP}}{3.5pt}}^{T}} \label{eq:covSSnewMAP}\,,
\end{align}
where both $\vb*{\widehat{B}_{\scaleto{\mathrm{MAP}}{3.5pt}}}$ -- rewritten explicitly as found in the derivation -- and $\vb*{\widehat{\mu_{s|r}^{\scaleto{\mathrm{MAP}}{3.5pt}}}}$ --intervening in $\vb*{\widehat{\Sigma}_{s|r}}=1/n_s\sum_{j=1}^{n_s}\left(\vb*{s}_j - \vb*{\widehat{\mu_{s|r}^{\scaleto{\mathrm{MAP}}{3.5pt}}}}\right)\left(\vb*{s}_j - \vb*{\widehat{\mu_{s|r}^{\scaleto{\mathrm{MAP}}{3.5pt}}}}\right)^{\vb*{T}}$-- estimators are identical to section \ref{sec:regML}. And we have written $\vb*{\Psi_{s|r}} = \vb*{\Psi_{ss}} - \vb*{\Psi_{sr}}\vb*{\Psi_{rr}}^{-1}\vb*{\Psi_{rs}}$. We have dropped the $P$ notation here since the reparametrization of the likelihood and prior in terms of the regression matrices instead of $\vb*{\Sigma}$ makes $p_s$ and $p_r$ appear separately.
As expected, the MAP estimator for $\vb*{\Sigma_{ss}}$ in this approach differs from the one derived in section \ref{sec:regML} since  the prior is not parameterization-invariant\footnote{We know that for two random vectors $\vb*{x}$ and $\vb*{y}$ with $\vb*{y}= h(\vb*{x})$, if $h$ is differentiable, then for probability distributions $\mathcal{P}_y(\vb*{y}) = \mathcal{P}_x(\vb*{x})\times \det(\vb*{J}_{h^{-1}(\vb*{y})})$ where $\vb*{J}$ is the Jacobian matrix. So under a reparametrization, the two distributions have no reason to peak at the same coordinates.}.

\subsection{Choice of the prior parameter $\vb*{\Psi}$}\label{sec:prior}
How should we choose the form of the parameter matrix $\vb*{\Psi}$? From now on, we consider that the surrogate and simulation summary statistics have the same dimension $p_s=p_r$, as this will be the case in section \ref{sec:exp}. In the context of an Inverse-Wishart distribution, $\vb*{\Psi}$ must be a $2p_s \times 2p_s$ symmetric positive-definite matrix.

Two generic choices we will present in the following with 1) blocks that are proportional to the identity matrix (the "identity" prior) or 2) blocks that are diagonal matrices  (the "diagonal" prior). In both cases, the coefficients and covariances are estimated based on the simulation-surrogate pairs.
Readers familiar with shrinkage estimators may recognize these as popular shrinkage targets (other common targets appear in Table 2 from \citet{schafer2005}). We will find that $\vb*{\Psi}$ appears in our estimators in an analogous way. For other particular applications, more tailored choices are of course possible. This may be the case when an approximate theoretical model for the covariances is available. As we will see in the numerical experiments in section \ref{sec:exp}, even the choice of a "diagonal" prior performs well and avoids the overfitting observed in the ML estimator as long as $n_p$ is chosen using the simple procedure described in section \ref{sec:priorHyper}.
The "identity" prior demonstrated improvement over the sample covariance of simulations for a much higher $n_s$ than the "diagonal" one, thus we will only present in section \ref{sec:exp} computations with the "diagonal" prior. We briefly describe the priors below.

\subsubsection{"Identity" prior}\label{sec:idPrior}
A common form adopted as a target for shrinkage estimates of covariance matrices is the "identity" prior: the auto-covariance of simulations and surrogates are proportional to identity matrices and the cross-covariance a diagonal matrix such that the correlation in each bin equals to $\rho \sigma_s \sigma_r$, with $\rho \in [0,1[$.

\begin{equation}
\text{\mbox{\huge{$\vb*{\Psi}$}}}_{\text{id}}\text{\mbox{\huge{$\equiv$}}}
    \begin{pmatrix}\;
      \begin{matrix}
    \sigma_s^2 & &  \text{\huge0} \\
     & \ddots  &   \\
     \text{\huge0} &  &  \sigma_s^2
      \end{matrix}
  & \rvline &  
      \begin{matrix}
    \rho \sigma_r \sigma_s & &  \text{\huge0} \\
     & \ddots  &   \\
     \text{\huge0} &  & \rho \sigma_r \sigma_s
      \end{matrix}\\
\hline
  \huge{\vb*{\Psi_{sr}^T}} & \rvline &
  \begin{matrix}
\sigma_r^2 & &  \text{\huge0} \\
 & \ddots  &   \\
 \text{\huge0} &  &  \sigma_r^2
  \end{matrix}
\end{pmatrix}\label{eq:PsiDumb}
\end{equation}
We require $\rho < 1$ for this matrix to be positive definite since $\mathrm{det}(\vb*{\Psi}_{id})=\left(\sigma^2_s\sigma^2_r(1-\rho^{2}) \right)^{p_s}$. This choice of $\vb*{\Psi}$ is very simple but still serves as a "regularizer" of the estimators from section \ref{sec:MAP}. We adopt an empirical Bayes approach, where we estimate $\rho$ and the variances $\sigma^2_r$ and $\sigma^2_s$ directly from the simulation-surrogate pairs. 

The estimated variance of $y_i=s_i$ or $r_i$, $1 \leq i \leq p_s$, is $\sigma_{y_i}^2 = \frac{1}{n_s-1}\sum_{j=1}^{n_s} \left(y_{i,j} - \overline{y_j} \right)^2$ and the estimated covariance between $s_i$ and $r_i$ is $\rho_i \sigma_{r_i}\sigma_{s_i} = \frac{1}{n_s - 1} \sum_{j=1}^{n_s} \left(s_{i,j} - \overline{s_i} \right) \left(r_{i,j} - \overline{r_i}\right)$. In equation \eqref{eq:PsiDumb}, each of the parameters $\sigma_s$, $\sigma_r$ and $ \rho \sigma_r \sigma_s$ is the average of the $p_s$ corresponding quantities, indexed by $i$.

Our numerical experiments with dark matter clustering statistics strongly preferred the "diagonal" prior we discuss next. 

\subsubsection{"Diagonal" prior}\label{sec:empBayes}

A natural choice to regularize the Maximum-Likelihood estimate for the covariance with simulations and surrogates is to use the estimated diagonal elements of $\vb*{\Sigma_{ss}}$, $\vb*{\Sigma_{sr}}$ and $\vb*{\Sigma_{rr}}$.

\begin{equation}
\text{\mbox{\huge{$\vb*{\Psi}$}}}_{\text{emp}}\text{\mbox{\huge{$\equiv$}}}
    \begin{pmatrix}
  \begin{matrix}
\sigma_{s_{1}}^2 & &  \text{\huge0} \\
 & \ddots  &   \\
 \text{\huge0} &  &  \sigma_{s_{p_s}}^2
  \end{matrix}
  & \rvline &  \begin{matrix}
\rho_{1} \sigma_{r_{1}} \sigma_{s_{1}} & &  \text{\huge0} \\
 & \ddots  &   \\
 \text{\huge0} &  &  \rho_{p_s} \sigma_{r_{p_s}} \sigma_{s_{p_s}}
  \end{matrix}\\
\hline
  \huge{\vb*{\Psi_{sr}^T}} & \rvline &
  \begin{matrix}
\sigma_{r_{1}}^2 & &  \text{\huge0} \\
 & \ddots  &   \\
 \text{\huge0} &  &  \sigma_{r_{p_r}}^2
  \end{matrix}
\end{pmatrix}\label{eq:PsiEmpBayes}
\end{equation}
The computation of each $\sigma_{s_i}^2$, $\sigma_{r_i}^2$ and $\rho_i \sigma_{s_i} \sigma_{r_i}$ is the same as from the "identity" prior above.

While having a very simple structure, we can see this prior as a more adapted correction of the eigenvalues of the block matrix $\vb*{\Sigma}$ based on the data, whereas $\vb*{\Psi}_{\text{id}}$ adds the same amount of correction on all the eigenvalues, regardless of the statistics at hand.

\subsection{(Cross-)validation to choose the prior hyperparameter $n_p$}\label{sec:priorHyper}
The hyperparameter $\nu$ (through $n_p= \nu + p_s + p_r + 1$) in equation \eqref{eq:invWishart} will be seen to determine the weight attributed to the prior in the closed-form solutions for $\vb*{\widehat{\Sigma}_{ss}^{\scaleto{\mathrm{MAP}}{3.5pt}}}$. For different statistics, and in terms of the maximum number of simulations $n_s$ one is able to run, varying $n_p$ via $\nu$ can significantly impact the quality of the covariance, as we will discuss in section \ref{sec:exp}. \\
We propose retaining a  small set $\left\{ \vb*{s}_{test} \right\}$ of test simulations such that $n_s = n_s^{\text{cov}} + n_s^{\text{test}}$, where $n_s^{\text{cov}}$ plays the role of the $n_s$ of the paired set in equations \eqref{eq:SigmassDelta} to \eqref{eq:covSSnewMAP}.

Consider the estimate $\vb*{\widehat{\Sigma}_{ss}^{\scaleto{\mathrm{MAP}}{3.5pt}}}(n_p)$ as a function of $n_p$. This can be computed with the same $n_s$ simulations, $\vb*{\Psi}$ prior and $n_r$ surrogates. Then an optimal $n_p$ can be computed by evaluating the MVN likelihood $\mathcal{L}(\{\vb*{s}_{test}\}| \vb*{\Sigma_{ss}}(n_p))$ which plays the role of a utility function. We find the $n_p$ that maximizes the likelihood on the test data\footnote{We compared this to using $K$-fold cross-validation but found no significant impact on the determination of the optimal $n_p$ comparatively to just evaluating the likelihood once without splitting the data.}

In our tests, we allow $n_p \in \llbracket 1, 4*p_s + 1 \rrbracket$, the upper bound being the smallest integer for which the Inverse-Wishart distribution is normalizable. While low $n_p$ values correspond to an improper prior we find in our numerical experiments that  the likelihood rises quickly for small values of $n_p$, with corresponding improvements to the MAP covariance estimates. Then a plateau is reached, with a shallow peak or plateau and a slow decrease as $n_p$ increases. Within the shallow peak the 
covariance estimates are robust to the precise value of $n_p$ and we advise choosing small values once the shallow regions is reached. We interpret this preference for low values as being due to the fact that for the simple, generic priors we used (block covariance with diagonal blocks, see section \ref{sec:prior}) and for the summary statistics at hand  a minimum of regularization by the prior is nearly optimal when $n_s$ is small. If specifically motivated prior matrices are available larger $n_p$ could perhaps become advantageous.

We present a summary of the estimation process, for the case of the block covariance estimation of section \ref{sec:regML}, in Algorithm \ref{algo}.
We obtained nearly identical results  treating $n_p$ as a hyperparameter and introducing a (Jeffreys) scale prior for it before maximization.

\begin{algorithm}
    \SetKwFunction{isOddNumber}{isOddNumber}
    \SetKwInput{KwIn}{Input}
    \SetKwInput{KwOut}{Output}
    % \SetKwInOut{KwIn}{Input}
    % \SetKwInOut{KwOut}{Output}
    
    \KwIn{A collection $\left\{ \vb*{x}_i \equiv \left( \vb*{s}_i,\vb*{r}_i\right) \right\}, i \in \llbracket 1,n_s \rrbracket$ of paired simulation and surrogate statistics; a large number of unpaired surrogate samples $\left\{ \rstar_j \right\}, j \in \llbracket 1,n_r \rrbracket$; a small number $n_s^{test}$ of simulation statistics; a  block prior $\vb*{\Psi}$; a set $\mathcal{N}_p$ of "prior weights" $n_p^k, k \leq card(\mathcal{N}_p)$.}
    
    % \KwOut{Covariance $\vb*{\widehat{\Sigma}_{ss}^{\scaleto{\mathrm{MAP}}{3.5pt}}}$ of the simulation summary statistics; simulation mean $\vb*{\widehat{\mu_{s|r}^{\scaleto{\mathrm{MAP}}{3.5pt}}}}$ given the surrogate; a likelihood evaluation of the sensitivity of the covariance estimate to $n_p$ for a fixed $n_s$.}
    \tcc{Here we compute the "loss" on a single test simulations set for simplification, but $K$-fold cross-validation is also an option.}
    \For{$n_p^k \in \mathcal{N}_p$}{
    Compute $\vb*{\widehat{\Sigma}_{ss}^{\scaleto{\mathrm{MAP}}{3.5pt}}}(n_p^k)$ using equations \eqref{eq:covRRMAP} to \eqref{eq:covSSMAP}.\\
    Compute the MVN likelihood $\mathcal{L}\left( \{\vb*{s}\}_{test}| \vb*{\Sigma_{ss}}(n_p) \right)$.
    }
    Determine $n_p^{\star}= \argmax_{\mathcal{N}_p} \mathcal{L}\left( \{\vb*{s}\}_{test}| \vb*{\Sigma_{ss}}(n_p) \right)$\\
    \KwRet{$\vb*{\widehat{\Sigma}_{ss}^{\scaleto{\mathrm{MAP}}{3.5pt}}}(n_p^{\star})$; $\vb*{\widehat{\mu_{s|r}^{\scaleto{\mathrm{MAP}}{3.5pt}}}}(n_p^{\star})$ }
    \caption{MAP Estimator for $\vb*{\Sigma_{ss}}$ given $n_p$}\label{algo}
\end{algorithm}

\subsection{Correction factor for the precision}

To compute confidence bounds of the cosmological parameters in the context of a likelihood-analysis, we need to invert the covariance matrix estimate. We briefly explain the correction used for the standard sample covariance.

\subsubsection{Classical result for the sample covariance}\label{sec:Hartlap}
It is well-known that taking the inverse of the bias-corrected version of the Maximum-Likelihood estimator from equation \eqref{eq:covSSML}, i.e $\gamma \vb*{\widehat{\Sigma}_{ss}}$ where $\gamma \equiv n_s/(ns - 1)$, results in a biased estimator of the precision matrix \citep{Kaufman1967, 2007A&A...464..399H}. For data sampled from a MVN, an unbiased estimator of the precision is

\begin{equation}
    \vb*{\widehat{P}_{ss}} = \frac{n_s - p_s - 2}{n_s -1}  \vb*{\widehat{\Sigma}_{ss}}^{-1}
\end{equation}

We chose, for this study, to include what in the cosmology literature is  referred to as the "Hartlap factor" to the inverse of the bias-corrected sample covariance of simulations summary statistics (including the truth using 15,000 simulations).

\subsubsection{For the MAP estimates}\label{sec:ourHartlap}

Our MAP estimates derived in section \ref{sec:regML} is constructed to  ensure that the result will be a symmetric positive semi-definite matrix. As a consequence, we lose formal unbiasedness but gain dramatically improved estimates according to multiple criteria, as discussed in section \ref{sec:exp}. If unbiasedness of the covariance estimate is important the method in \carpcov
can be used.

\section{Numerical Experiments on $\Lambda$CDM simulation statistics}\label{sec:exp}

\subsection{Simulation and surrogate data}
The  simulation and surrogate solvers we use are identical to those \carpcov\ and \carpool. We recall the main points here for convenience. For more details please refer to \carpool. 
The solvers evolve $\mathcal{N}_\mathrm{p} = 512^3$ Cold Dark Matter (CDM) particles in a box volume of $\left( 1000~h^{-1} {\rm Mpc} \right)^3$.
The simulation-surrogate sample pairs take the same Second-order Lagrangian perturbation theory (2LPT) initial conditions at starting redshift $z_{i}=127.0$.

\subsubsection{$N$-body solver}

We downloaded the $N$-body snapshots clustering statistics from the publicly available \textit{Quijote} simulation suite\footnote{ \url{https://quijote-simulations.readthedocs.io/en/latest/}} \citep{2020ApJS..250....2V}. The solver for all the simulations is the TreePM code \texttt{GADGET-III} built upon the previous version \texttt{GADGET-II} by \citet{2005MNRAS.364.1105S}. The force mesh grid size to solve the comoving Poisson equation at each timestep is $\mathcal{N}_\mathrm{m} = 1024$. In the following, we will use the sample covariance of all 15,000 available realizations of the fiducial cosmology as the simulation "truth", or more precisely the best covariance estimate we have access to.

\subsubsection{Surrogate solver}

We generate the fast surrogate samples with The COmoving Lagrangian Acceleration (COLA) method from \citet{Tassev_2013} (see also \citet{2020A&A...639A..91L}), which allows generating approximate gravitational $N$-body outputs using a smaller number of timesteps than our simulation code.
The principle of COLA is to add residual displacements, computed with a Particle-Mesh (PM) $N$-body solver, to the trajectory given by analytical LPT approximations (usually first- or second-order). \citet{Izard_2016} proposed tests of the accuracy and computational cost of COLA against $N$-body simulations at different redshifts and with different timestepping parameters.
Like in \carpool\ and \carpcov, we used the parallel MPI implementation \texttt{L-PICOLA} developed by \citet{Howlett_2015}, with a coarser force mesh grid size of $N_m^{\mathrm{cola}}=512$.

\subsubsection{Post-processing of snapshots}

To extract the summary statistics from our \texttt{L-PICOLA} snapshots, we used the exact same code modules and parameters used to compute the clustering statistics available in the \textit{Quijote} data outputs. Therefore, the simulation and surrogate summary statistics have the same dimension $p_s = p_r$. 
We transform the snapshots in density contrast fields with the Cloud-In-Cell (CiC) algorithm.
For the matter power spectra and the correlation functions, we used the \texttt{Python3} module \texttt{Pylians3} \footnote{\url{https://github.com/franciscovillaescusa/Pylians3}}, For the bispectra, the results of which appearing in Appendix \ref{app:Bk}, the post-processing code is \texttt{pySpectrum}\footnote{Available at \url{https://github.com/changhoonhahn/pySpectrum}}. More details can be found in \carpool.

\subsection{The CARPool Bayes estimator and results on clustering statistics}

The following tests of the Bayesian covariance estimation approach in this paper use the sample covariance matrix with all the simulations we have ($n_s^{truth}=15,000$) as the "truth" to compare with other estimates. Within the main part of this paper we only present a subset of the estimators that gave the best match in terms of parameters constraints with respect to the truth.

In particular, we use the MAP estimator from section \ref{sec:regML} with the "diagonal" empirical Bayes prior $\vb*{\Psi}_{\text{emp}}$, equation \eqref{eq:PsiEmpBayes}, estimated on the paired set of $n_s^{cov}$ simulations and surrogates. All our MAP covariance estimates with simulations and surrogates use the optimal $n_p$ determined through the process described in Algorithm \ref{algo} with a small number of test simulations. We display the total number of simulations used for each covariance matrix estimate as $n_s = n_s^{\text{cov}} + n_s^{\text{test}}$.

In the following, we will refer to this approach as the "CARPool Bayes" estimator.

We find that the alternative  estimator written in terms of the regression parameters, section \ref{sec:newMAP}, performs  comparatively poorer than the CARPool Bayes estimator.  We show an example on the power spectrum covariance and discuss the reasons for this in Appendix \ref{sec:mapRegExp}. Briefly summarized, this estimator requires using a proper prior and therefore affords us less flexibility in choosing the weight of the prior. It therefore tends to give covariance estimates that are more sensitive to the choice of the prior parameters.

The plan for the remainder of this sections is as follows: we will first present the  power spectrum results in more details that were already partially described in section \ref{sec:illustration}. 

Then, we turn to the real space $2$-point correlation function. This is an interesting case because it illustrates the power of limiting the range of the estimator to the set of all positive definite covariance matrices, a feature of the Bayesian version of CARPool. The unbiased CARPool approach to the covariance matrix in \carpcov\ failed to yield positive-definite covariance estimates for this application in spite of a significant reduction of variance for the covariance matrix individual elements.

For a complete comparison with \carpcov, we also computed results on the bispectrum covariance matrix. Since these show similar, large improvement over the \carpcov\ approach as for the power spectrum, we relegate details to Appendix \ref{app:Bk}.

\subsubsection{Matter power spectrum covariance}

The matter power spectrum $[{Mpc^{3}}]$, at wave number $k [h {\rm Mpc^{-1}}]$, under the conditions of homogeneity and isotropy (\textit{cosmological principle}), is the average in 3D Fourier space of $|\delta\left(k\right)|^2, k \in \left[ k-\Delta k/2, k+\Delta k/2\right]$, where $\delta(\boldsymbol{x})$ is the matter density contrast in real space. For each snapshot, we compute $\delta(\boldsymbol{x})$ on a square grid of size $1024$ with the Cloud-In-Cell (CIC) algorithm.  
The following analysis is for $k \in \left[\num{8.900e-3}, 1.0 \right] h {\rm Mpc^{-1}}$. We have then $p_s=158$ linearly space bins. Note that the power spectrum is not compressed unlike in \carpool\ and \carpcov, making the covariance estimation tasks more difficult.

Figure \ref{fig:fisherPk}  shows that using only $n_s= 10 + 5$ simulations with paired surrogates and an additional set of surrogate samples, we get confidence bounds for the cosmological parameters which are very close to the ones given by the "true" sample covariance using 15,000 simulations (for $p_s=158$). This result is all the more encouraging that with only $10$ simulations we would get a sample covariance of rank at most $9$. In other words, we can see the small set of simulations in the "CARPool Bayes " estimate a correction to the eigenvalues and eigenvectors of the precision matrix computed from a biased but correlated surrogate. We also show in Appendix \ref{app:morePk} the relatively small gain, in terms of closeness of the parameters confidence contours to the truth, of running $n_s = 40 + 10$ simulations for comparison.

\begin{figure}
    \includegraphics[width=\columnwidth]{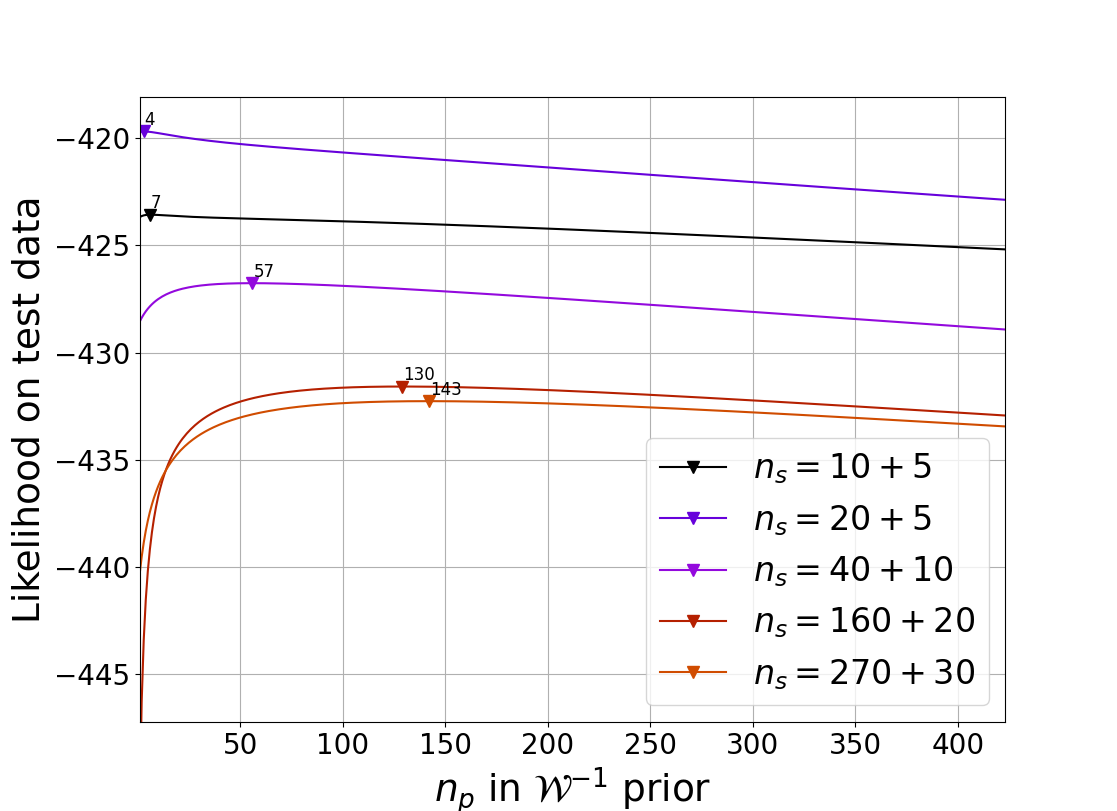}
    \caption{Illustration of step $5$ of Algorithm \ref{algo} for the matter power spectrum example: for fixed $n_s = n_s^\text{cov} + n_s^\text{test}$ and fixed prior $\vb*{\Psi}_{\text{emp}}$, we compute the likelihood of $\vb*{\widehat{\Sigma}_{ss}^{\scaleto{\mathrm{MAP}}{3.5pt}}}(n_p)$ on test simulation samples.}
    \label{fig:nshrinkPk}
\end{figure}

Here we examine the procedure to determine the best $n_p$ for a given $n_s$ and $\vb*{\Psi}$ in Figure \ref{fig:nshrinkPk}. There are several points to notice here:
\begin{enumerate}
\item For $n_p \approx 1$, especially for $n_s \geq p_s + 1$ (when the sample covariance can be full-rank), the likelihood on test data rapidly increases. It shows for this case that a minimum of "regularization" brought by the prior greatly improves the estimate of $\vb*{\Sigma_{ss}}$.
\item Around the empirical $n_p^{\star}$, the likelihood is rather flat and slowly decreases when $n_p > n_p^{\star}$. In other words, once a certain threshold of "improvement" is reached with $n_p$, misestimating $n_p$ does not radically worsen the estimate of $\vb*{\Sigma_{ss}}$.
%\item Using a different small test set gives a different $n_p^{\star}$ for fixed $n_s$, as the $n_s = 10 + 5$ and $n_s = 10 + 10$ cases illustrate. Given the previous remark, we saw no noticeable difference in terms of Fisher confidence bound in the end, therefore we choose the configuration with the smallest total number of simulations.
\end{enumerate}

\begin{figure*}
    \includegraphics[width=\textwidth]{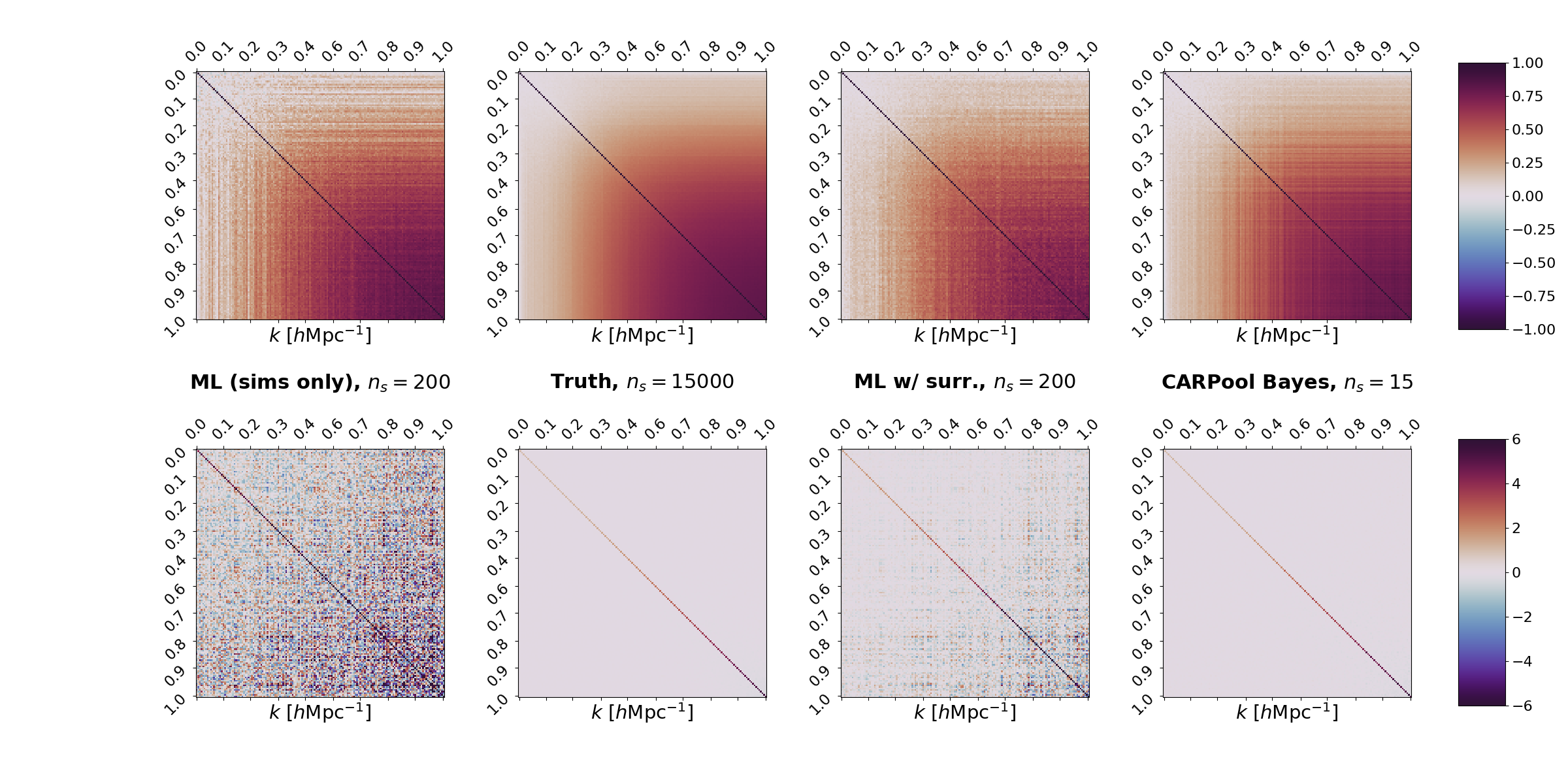}
    \caption{Matter power spectrum covariance estimates (top row) and their inverse (bottom row). We show the covariances as correlation matrices with the normalization $\vb*{D}
    ^{-1}\vb*{\widehat{\Sigma}}\vb*{D}
    ^{-1}$ with the diagonal $\vb*{D} = \sqrt{\mathrm{diag}\left( {\vb*{\widehat{\Sigma}}}\right)}$, and the precision matrices below are the inverses of these correlation matrices. Columns from left to right show the standard sample covariance estimate from 200 simulations; the reference covariance from 15,000 simulations; the Maximum Likelihood estimate using the  combination of surrogates and 200 simulations (section \ref{sec:ML}); and the "CARPool Bayes" estimate with a "diagonal" prior, combining $n_s = 10 + 5$ \texttt{GADGET simulations} with surrogates.}
    \label{fig:visualPk}
\end{figure*}

In Figure \ref{fig:visualPk}, we visualize the estimated covariance matrices (top row) and their inverse (bottom row). For the "CARPool Bayes" estimate with the prior $\vb*{\Psi}_{\text{emp}}$, i.e. our "headline" estimate with $n_s = 10 + 5$ that gives the confidence bounds in Figure \ref{fig:fisherPk}, we notice some structure in the covariance due to the small number of simulations. The closeness to the truth of the "CARPool Bayes" covariance with very few simulations is particularly visible for the structure of the precision matrix. It can be seen that at low $k$, where the correlation between surrogates and simulations is particularly high, the CARPool Bayes estimate (and the Maximum Likelihood estimate without the prior) is significantly less noisy than the  standard estimator even though it uses an order of magnitude less simulations.

\begin{figure}
    \includegraphics[width=\columnwidth]{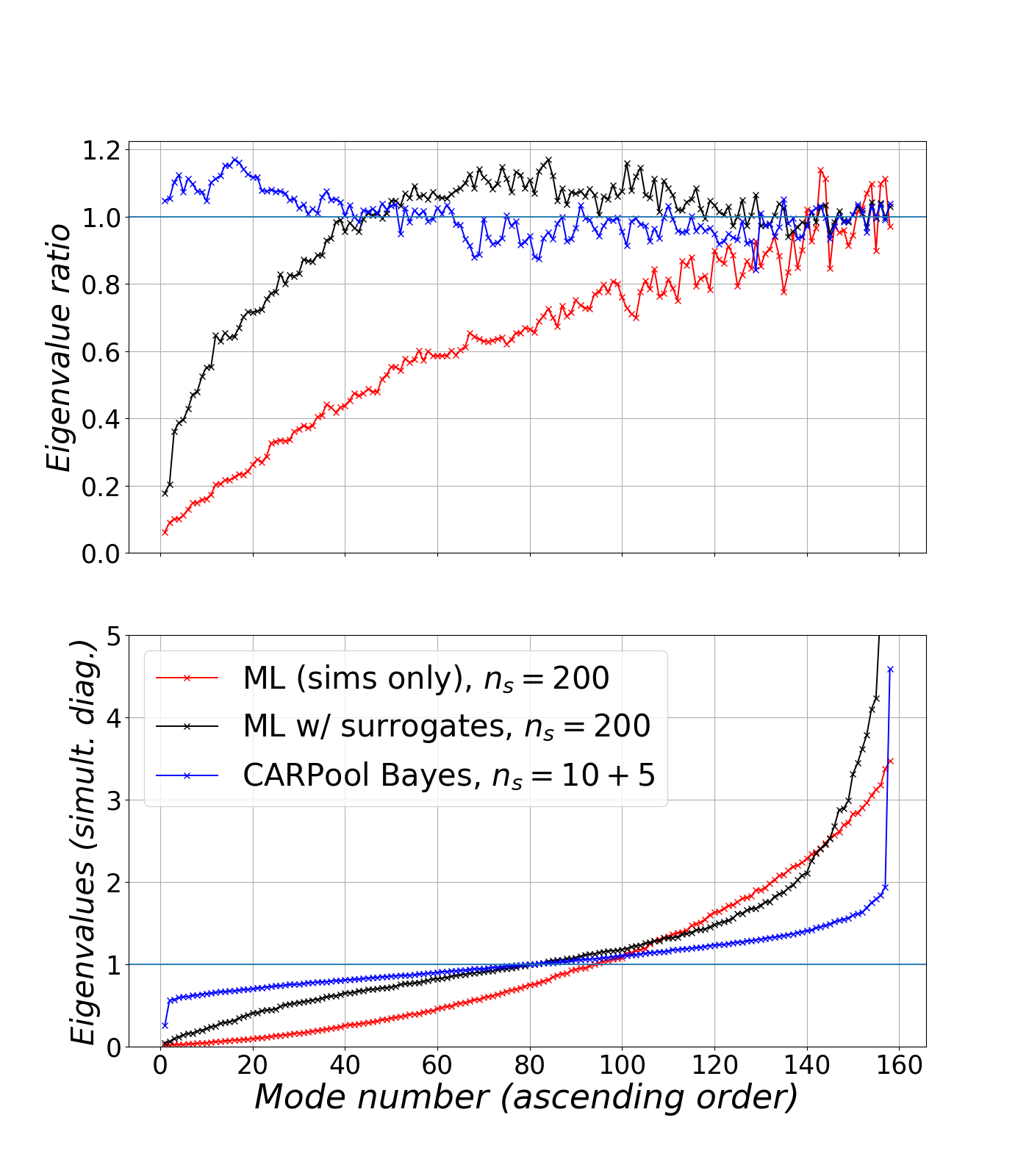}
    \caption{Comparison of the CARPool Bayes covariance estimate (section \ref{sec:regML}),  the standard ML estimator, and the ML estimator combining simulations and surrogates (section \ref{sec:ML}) with the large-sample "truth" in the spectral domain. We show ordered eigenvalue ratio at the top and co-diagonalisation coefficients at the bottom. The CARPool Bayes estimator avoids the characteristic underestimation of small eigenvalues for covariance matrices estimated from a small number of samples. See discussion in the text. }
    \label{fig:eigenPk}
\end{figure}

In Figure \ref{fig:eigenPk}, we compare the covariance estimates to the large-sample "truth" in the spectral domain, showing the eigenvalues and the co-diagonalization coefficients 
\footnote{For $\vb*{A}$ and $\vb*{B}$ two $p \times p$ real symmetric matrices, if $\vb*{A}$ is positive definite, then there exists a matrix $\vb*{M}$ such that $\vb*{M^TAM} = \vb*{I}_p$ and $\vb*{M^TBM} = \vb*{D}$ with $\vb*{D}= \text{diag}(d_1, \dots, d_p)$. We call the $d_i$ "co-diagonalization coefficients."  This is a simplified statement from theorem A9.9 in \citet{Muirhead}. If $
\vb{D}=\unit_{p}$ then $\vb{A}=\vb{B}$.}. 

At the top, we show the ordered eigenvalues ratio of each matrix. A ratio far from $1$, and especially close to zero for the smallest eigenvalues as for the standard sample covariance, indicates a very poor conditioning of the matrix. 
At the bottom, we see the co-diagonalization coefficients.
A horizontal line at $1$ would indicate that the matrices are identical. 
The CARPool Bayes estimate clearly outperforms the other estimates and avoids the characteristic underestimation of small eigenvalues for covariance matrices estimated from a small number of samples.

\subsubsection{Matter correlation function covariance}

The example of the two-point matter correlation function $\xi(\boldsymbol{r})$ for $\boldsymbol{r} \in \left[5.0, 160.0\right] h^{-1} {\rm Mpc}$ ($p_s=159$) is of particular interest in our study. With the variance reduction approach in \carpcov\ for the covariance matrix, we found no improvement over the standard estimator. While unbiased and strongly reducing the errors of all individual elements of the covariance matrix  the resulting matrix  failed to be positive-definite. This means that no estimate for the precision matrix could be obtained, as would be required to derive Fisher matrices or for a likelihood approximation to derive parameter constraints.

\begin{figure*}
    \includegraphics[width=\textwidth]{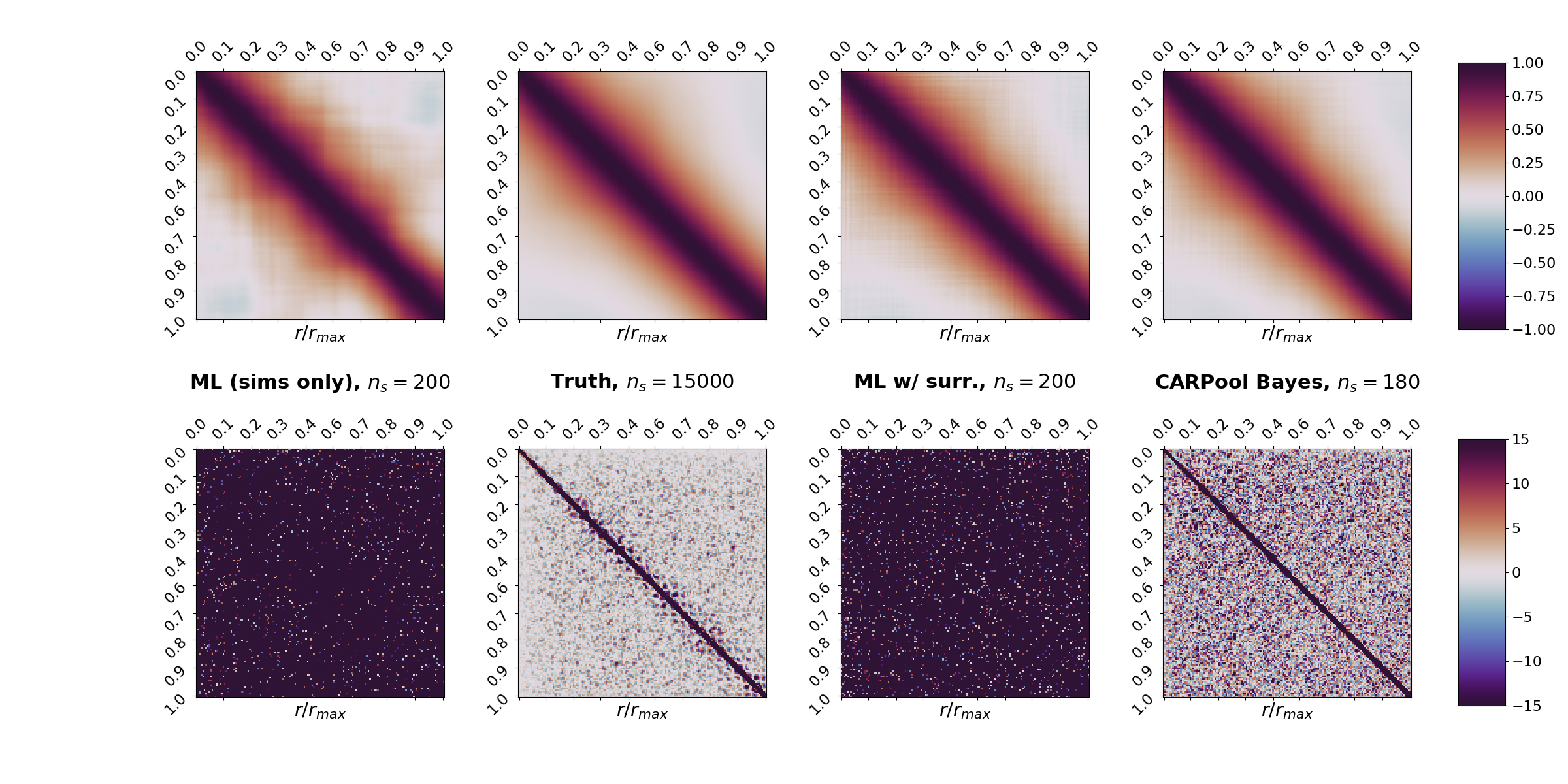}
    \caption{Matter correlation function covariance estimates (top) and their inverse (bottom), shown similarly to Figure \ref{fig:visualPk}. The "CARPool Bayes" estimates uses $n_s = 160 + 20$ \texttt{GADGET simulations}.}
    \label{fig:visualCorr}
\end{figure*}

As we can observe in Figure \ref{fig:visualCorr}, the structure of the covariance is particular, with a band of high-magnitude covariances around the diagonal of variances. As a result, the precision estimate based on the standard sample covariance estimator is very noisy for $n_s=200$, which we compare with our estimate including surrogates from Algorithm \ref{algo}, with $n_s = 160 + 20$. Looking at the precision matrices (bottom row) would indicate a significantly better recovery of the structure of the true precision.

\begin{figure*}
    \includegraphics[width=\textwidth]{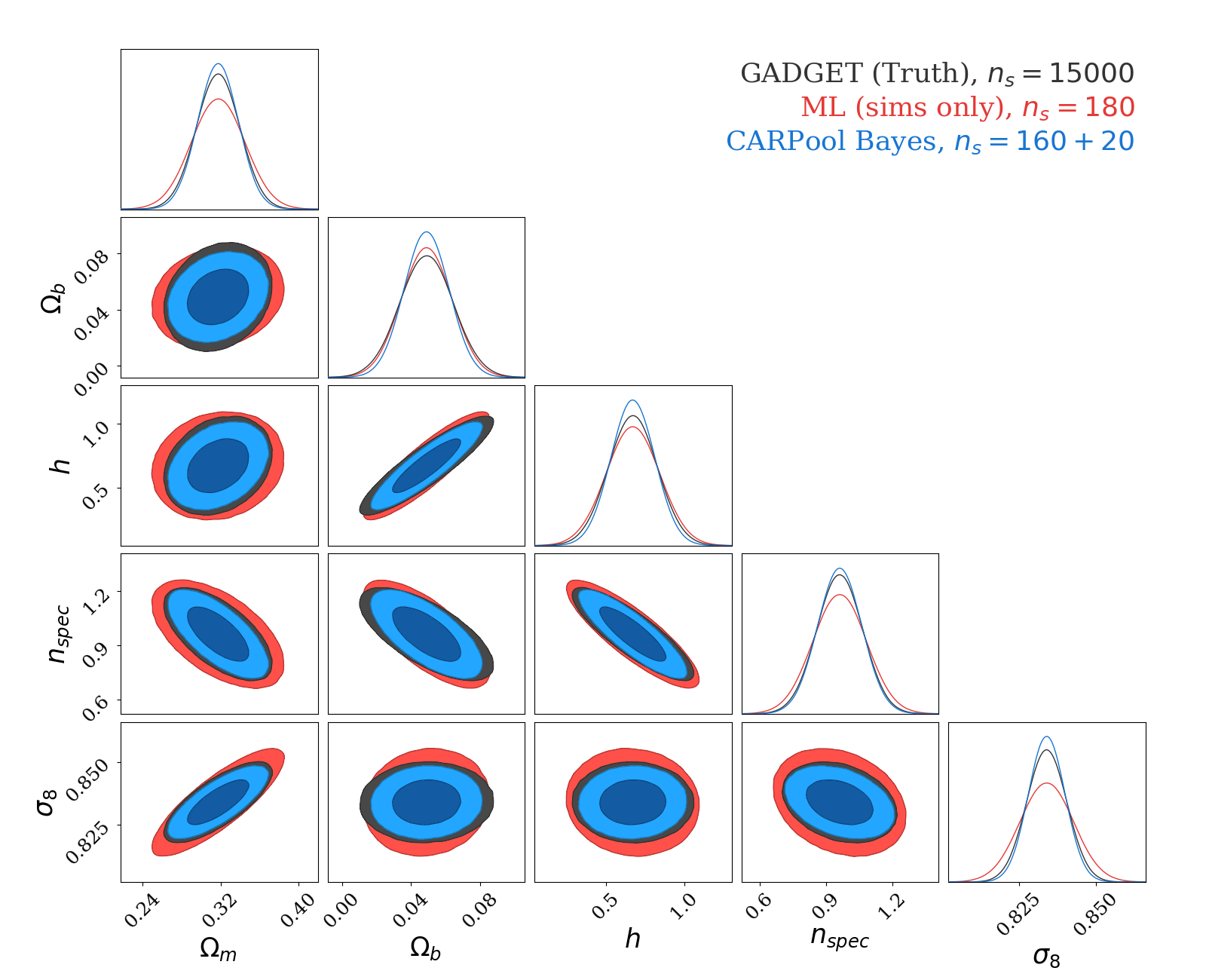}
    \caption{Confidence contours of the cosmological parameters computed using the Fisher matrix based on the estimated matter correlation function covariance matrix. The estimators which we compare are the same as in Figure \ref{fig:fisherPk}.}
    \label{fig:fisherCorr}
\end{figure*}

In terms of cosmological parameter forecast constraints, as shown in Figure \ref{fig:fisherCorr} , we get a slight improvement with respect to the sample covariance matrix (and the precision including the Hartlap factor), but not nearly as large as for the matter power spectrum.
%Either way the improvement over the sample covariance is debatable, but the point here is that we get "something" contrarily to \carpcov}.
The CARPool Bayes estimate with $n_s = 160 + 20$ produces  bounds for $\Omega_m$, $n_s$ and $\sigma_8$ that are closer to the truth than with the sample covariance with $n_s = 180$. But the confidence regions for $\Omega_b$ and $h$ are not improved.

\begin{figure}
    \includegraphics[width=\columnwidth]{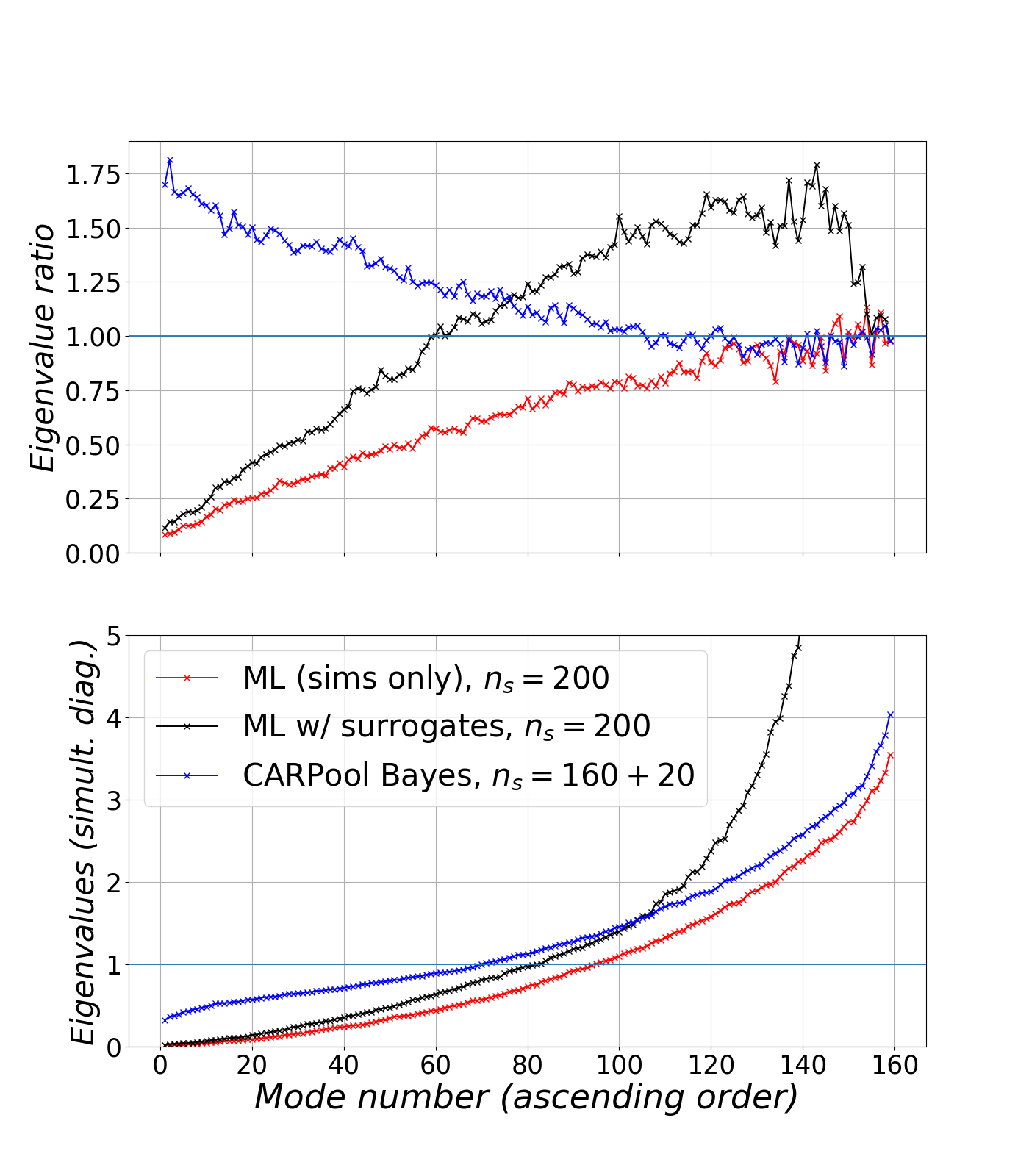}
    \caption{Same  as Figure \ref{fig:eigenPk} for the matter correlation function.}
    \label{fig:eigenCorr}
\end{figure}

Similarly to the previous section, in Figure \ref{fig:eigenCorr}, the "CARPool Bayes" estimator raises up the smallest eigenvalues -- as well as the smallest "co-diagonalization" coefficient -- contrarily to the ML solutions with and without surrogates  where they are close to $0$.

\begin{figure}
    \includegraphics[width=\columnwidth]{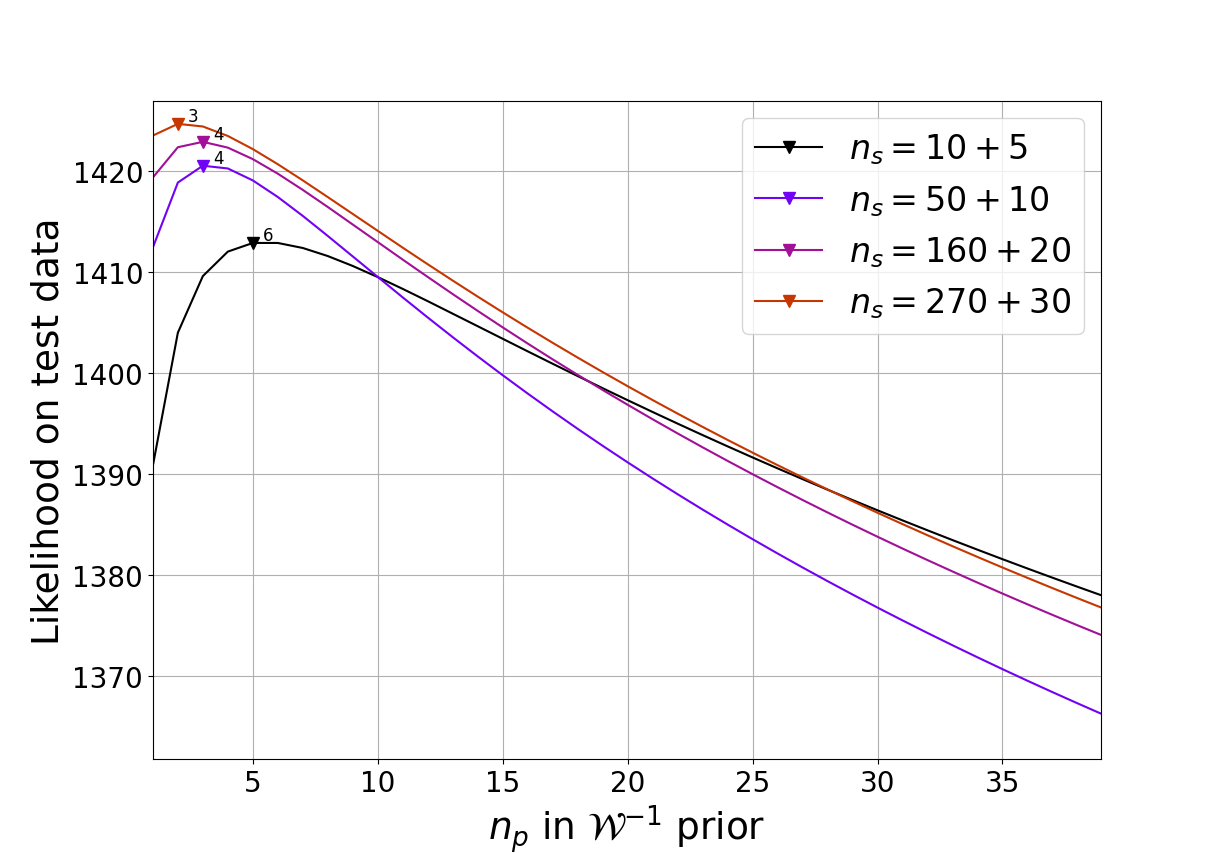}
    \caption{Same plot as in Figure \ref{fig:nshrinkPk} for the matter correlation function, still with $\vb*{\Psi}_{\text{emp}}$.}
    \label{fig:nshrinkCorr}
\end{figure}

The wide band of correlations visible in Figure \ref{fig:visualCorr} indicate that our choice of "diagonal" prior is not optimal for this case. Choosing a prior with a more gradual falloff of correlation from the diagonal would likely produce better results.
Figure \ref{fig:nshrinkCorr} indicates that for various number of simulations $n_s$, the CARPool Bayes estimator for the matter correlation function covariance consistently prefers low $n_p$ (i.e. prior weight) values with the "diagonal" prior from section \ref{sec:empBayes}.

In summary, the application to the matter correlation function, demonstrates that the CARPool Bayes estimator is guaranteed to produce positive definite matrices. It visibly improves the structure of the precision matrix (Figure \ref{fig:visualCorr}) and the relative errors of the small eigenvalues (Figure \ref{fig:eigenCorr}). This translates into some, but not all, parameter confidence bounds being closer to the truth than for the sample covariance based on $180$ simulations.

\section{Discussion and conclusion}\label{sec:conclusion}
We consider the problem of estimating the covariance matrices of cosmological summary statistics within a Bayesian framework, when paired simulations and surrogates are available.

This study constitutes an extension of the CARPool principle, presented in \carpool\ and applied to covariance matrices in \carpcov. Our method improves on the latter work by solving a Maximum A Posterior optimization directly in the space of symmetric positive semi-definite matrices and allows introducing priors in analogy to frequentist shrinkage estimators. We prove that our approach, dubbed CARPool Bayes,  guarantees positive definite estimates, for the price of abandoning the guarantee of unbiasedness of individual covariance matrix elements provided by the first order estimator described in \carpcov.

By casting \carpool\ in a Bayesian framework we provide a new solution to covariance estimation with simulations and surrogates. We demonstrate that this estimator can strongly improve over previous approaches such as the sample covariance or the first-order CARPool approach in \carpcov\ according to multiple criteria. These improvements are particularly noticeable for the inverse covariance or precision matrix required for many applications such as computing Fisher matrices, or for the Gaussian likelihood approximations frequently used for parameter estimation. 

Our Bayesian approach combines estimations for the both the mean (through the well known regression $\vb*{\mu_{s|r}}$, equation \ref{eq:muSRMAP}) and the covariance of simulation summary statistics using surrogates. In this paper we focused on showing the results for  the simulation covariance estimates $\vb*{\Sigma_{ss}}$ since this is the first time that the control variate approach  has been cast in a Bayesian framework for covariance estimation.

Our Bayesian approach used a multivariate Gaussian model for the simulations and surrogates and includes a conjugate Inverse-Wishart distribution prior for the covariance matrix. In the generic case we found a "diagonal" prior on the block covariance of simulation and surrogate summary statistics, whose diagonal elements were evaluated on simulation-surrogate pairs, section \ref{sec:empBayes}, to give excellent results, especially for the matter power spectrum and  bispectrum. We obtain the same confidence bounds as with the true covariance of the power spectrum with $p_s = 158$ bins up to  $k \approx 1.0$ $h {\rm Mpc^{-1}}$ with only $n_s = 10 + 5$  simulations. In this case, we can think of the actual $10$ simulations of the covariance estimate as correcting the eigenspectrum of the well-converged covariance of the correlated surrogate that incorporates many samples. 

The same outstanding gain appears for  the bispectrum,  as we show in Appendix \ref{app:Bk}  for two triangle configurations. This demonstrates the superiority of the CARPool Bayes approach over \carpcov.

Regarding the $2$-point matter correlation function in real space, we do get positive-definite estimates by construction---this is not guaranteed in \carpcov---and we obtain a slight improvement on the parameter constraints with respect to the sample covariance of simulations when $n_s \gtrsim p_s + 1$ is close to the dimension of the summary statistics. But in a case where running a high enough number (we tested  $n_s=300$) of simulations is possible, the gain over the sample covariance diminishes as $n_s$ increases, at least regarding the impact of the matter correlation function covariance  on the parameter constraints.

Throughout this study, we applied the  "diagonal" empirical Bayes prior through taking the diagonal of each block of the summary statistics as a concatenation of the simulation and surrogate output. Using the former was sufficient to demonstrate the capability of the method for the case where we consider the problem of estimating the whole block covariance $\vb*{\Sigma}$ to then extract the simulation block $\vb*{\Sigma_{ss}}$. We derived new MAP estimators in section \ref{sec:newMAP} where we directly estimate the regression parameters allowing to compute the simulation block of the covariance using $ \vb*{\Sigma}_{ss} \equiv \left( \vb*{\Sigma}/\vb*{\Sigma}_{rr} \right)+\vb*{\Sigma}_{sr}\vb*{\Sigma}_{rr}^{-1}\vb*{\Sigma}_{rs}$ from the hypothesis of data sampled from a MVN distribution. This new estimator did not provide improvement in our tests over the sample covariance, which we attribute to the strong prior dependence inherent in it. We leave for future studies the question whether the this different parametrization can turn beneficial for cosmological survey forecasts when  theoretically motivated covariances for the prior are available.

\subsection{Generating samples from the posterior}
As an alternative to focusing on closed-form point-estimates of the covariance by taking the Maximum A Posterior (MAP) of the posterior distribution we could have considered generating samples of the simulation covariance matrix  from the posterior. This is possible using a Gibbs sampling approach where we explicitly include the missing simulations $\sstar$ as latent variables. We briefly sketch the approach here: first draw $\vb*{\Sigma}$  from a conditional Inverse-Wishart for positive (semi-)definite covariance matrices given the data augmented by the latest $\sstar$ sample. Since the augmented data is a complete set of simulation-surrogate pairs the $\vb*{\Sigma}$ sample would therefore be guaranteed to be positive (semi-)definite. Simply extracting the simulation auto-correlation block from $\vb*{\Sigma}$ would produce samples from the marginal posterior for $\vb*{\Sigma_{ss}}$.

While samples from the marginal posterior would potentially be useful to propagate the uncertainty in the estimates, or to study other posterior summaries such as the posterior mean,  we do not explore this approach further, for two reasons: one is computational cost though that is perhaps tolerable for summaries with moderate dimension (\textit{i.e.}, up to $\order{100}$); the other is that we would like to obtain a point estimate for the covariance that we can use in other contexts, without worrying if the Monte Carlo estimate, e.g., of the posterior mean of the signal covariance, has converged to sufficient accuracy.

\subsection{Potential for future applications in cosmology and beyond}
Our numerical experiments demonstrate the  capability of running fewer intensive simulations in order to get theoretical predictions of the means and covariances of observables for next-generation surveys. Many additional applications of these techniques remain to be explored.  The free choice of what to use as surrogates makes our methods very broadly applicable. 

Some surrogates might be useful because they are nearly free computationally. A case in point would be Eulerian linear perturbation theory for the power spectrum applied to the initial conditions of an $N$-body simulation. In this case each simulation comes with the paired surrogate for free (since the initial conditions are necessary to run the simulation in the first place) and its expectation and covariance can be computed analytically nearly for free. It could be argued that such automatic surrogates ought to be exploited systematically when predicting commonly used clustering statistics from simulations. A very similar application of this idea to a non-perturbative statistic would be to the computation of halo number functions: apply the Press-Schechter approach to the initial conditions as a surrogate for the mass function for a given realisation. In this example, the classical Press-Schechter formula provides the expectation of the surrogate and would reduce the variance in the number function for the largest (and rarest) clusters in the simulations, thus increasing the effective volume of the simulations.

In other cases, the surrogates may consist of costly simulations that have already been run at a different set of parameters. In this case it may be possible to "update" the means and covariances from the previous simulation set to a new set of parameters by pairing a small number of the existing old simulations (now surrogates) with the same number of new simulations. 

The availability of perturbative results and analytical estimates, the increasing need for accurate simulations to analyze current and upcoming data sets in all subfields of cosmology, and the vast parameter space to explore with cosmological simulations make it likely that the concepts described here will continue to find powerful applications. We look forward to seeing the cosmological advances that CARPool will enable.

\section*{Acknowledgements}
We warmly thank Ethan Anderes and Francisco Villaescusa-Navarro for stimulating discussion and feedback.
N.C. acknowledges funding from LabEx ENS-ICFP (PSL). B.D.W. acknowledges support by the ANR BIG4 project, grant ANR-16-CE23-0002 of the French Agence Nationale de la Recherche; and the Labex ILP (reference ANR-10-LABX-63) part of the Idex SUPER, and received financial state aid managed by the Agence Nationale de la Recherche, as part of the programme Investissements d'avenir under the reference ANR-11-IDEX-0004-02.
The Flatiron Institute is supported by the Simons Foundation.
This work has made use of the Infinity Cluster hosted by Institut d'Astrophysique de Paris.

\section*{Data availability}
The data samples underlying this article are available through \textit{globus.org}, and instructions to reproduce the summary statistics from snapshots can be found at \url{https://github.com/franciscovillaescusa/Quijote-simulations}. Additionally, a \texttt{Python3} package with code examples and documentation is provided at \url{https://github.com/CompiledAtBirth/pyCARPool} to experiment with CARPool.
\appendix

\section{Derivation of estimators and proof of positive definiteness using Expectation Maximization}\label{app:EM}

\subsection{Expectation Maximization}
In this section, we aim at showing the equivalence of the results given by the Expectation-Maximum algorithm -- which naturally comes to mind in the presence of missing samples -- and the simple result from the Maximum-Likelihood and Maximum a Posteriori problems formulated in sections \ref{sec:ML} and \ref{sec:regML}.

\subsubsection{Iterative algorithm}
The Expectation-Maximization (EM) algorithm \citep{10.2307/2984875} is an iterative technique  to  maximize the likelihood (or posterior) in the presence of missing data. Briefly, it works by casting the problem as a sequence of simpler optimization problems. Each iteration consists of two steps: the \textit{E-step} which removes the missing data from the log-likelihood by taking its expectation with respect to the missing data assuming the current iterates are the true values of the  parameters; and the \textit{M-step} which updates the parameters by finding their values that maximize the expected log-likelihood from the E-step. We focus in this appendix on the covariance estimation; including the solution for the estimators of the mean is straightforward and we give the result in the main text. 

We recall Eq.~\eqref{eq:LogLikelihood} here for convenience as a starting point
\begin{align}\label{eq:logLikelihood}
   &-2\lnb{\mathcal{L}(\{\vb*{x}\},\{\xstar\}| \vb*{\Sigma})}= (n_s+n_r)\lnb{ \det\left(2\pi \vb*{\Sigma}\right)} \nonumber\\
   &+\qty(\sum_{i=1}^{n_s} \vb*{x}_i^{\vb*{T}} {\vb*{\Sigma}}^{-1}\vb*{x}_i)
   +\qty(\sum_{j=1}^{n_r} \xstar_j{}^{\vb*{T}} \vb*{\Sigma}^{-1}\xstar_j)\,,
\end{align}

\paragraph*{E-step.}
Consider conditional expectation of the log-likelihood over missing data $\sstar$ given the (observed) data and the covariance at the $k$-th step, $\vb*{\Sigma}[k]$
\begin{align}
    &-2\mathbb{E}_{\sstar|\rstar}\qty[\lnb{\mathcal{L}(\{\vb*{x}\},\{\xstar\}| \vb*{\Sigma}[k])}]=(n_s+n_r)\lnb{ \det\left(\vb*{\Sigma}[k]\right)}\nonumber\\
&\qquad+\qty(\sum_{i=1}^{n_s} \vb*{x}_i^{\vb*{T}} {\vb*{\Sigma}[k]}^{-1}\vb*{x}_i)
+\mathbb{E}_{\sstar|\rstar}\qty[\qty(\sum_{j=1}^{n_r} \xstar_j{}^{\vb*{T}} \vb*{\Sigma}[k]^{-1}\xstar_j)] + c\nonumber
 \label{eq:Eloglike}
\end{align}

Using linearity of expectation we can look at each summand of the last term on the RHS
\begin{equation}
    \mathbb{E} \qty[\qty(\xstar{}^{\vb*{T}}_i \vb*{\Sigma}[k]^{-1} \xstar_i)]=\tr(\vb*{\Sigma}[k]^{-1}\mathbb{E}\qty[\xstar_i\xstar_i{}^{\vb*{T}}])
\end{equation}

Define
\begin{equation}
    \vb*{A}_{i} \equiv \mathbb{E}\qty[\xstar_i \xstar_i{}^{\vb*{T}}] = \mqty(\vb*{A}_{i,\vb*{ss}} & \vb*{A}_{i,\vb*{sr}} \\ \vb*{A}_{i,\vb*{sr}}^{\vb*{T}} &\vb*{A}_{i,\vb*{rr}})\,.
\end{equation}

Then 
\begin{align}
    \vb*{A}_{i,\vb*{ss}}&=(\vb*{\Sigma}/\vb*{\Sigma_{rr}})+\vb*{B}\rstar_i\rstar_i{}^{\vb*{T}}\vb*{B^T}\label{eq:Ass}\\
    \vb*{A}_{i,\vb*{sr}}&=\vb*{B}\rstar_i\rstar_i{}^{\vb*{T}}\label{eq:Asr}\\
    \vb*{A}_{i,\vb*{rr}}&=\rstar_i\rstar_i{}^{\vb*{T}}\label{eq:Arr}
\end{align}
We stress that equations \eqref{eq:Ass}, \eqref{eq:Asr} and \eqref{eq:Arr} depend on $k$ because we use $\vb*{\Sigma}[k]$ as $\vb*{\Sigma}$. 

Writing 
$$n_s \widehat{\vb*{\Sigma}}=\sum_{i=1}^{n_s}\vb*{x}_i\vb*{x}_i^{\vb*{T}}$$ and 
$$n_r \vb*{A}[k]=\sum_{i=1}^{n_r}\vb*{A}_i[k],$$ we find the expected log-likelihood
\begin{align}\label{eq:EloglikeFinal}
    &-2\mathbb{E}_{\sstar|\rstar}\qty[\lnb{\mathcal{L}(\{\vb*{x}\},\{\xstar\}| \vb*{\Sigma}[k])}]= (n_s+n_r)\lnb{\det(\vb*{\Sigma}[k])} \\
    &\qquad+\tr[\vb*{\Sigma}^{-1}\qty(n_s\widehat{\vb*{\Sigma}} +n_r\vb*{A}[k])]+ c\nonumber 
\end{align}

\paragraph*{M-step.}
Maximizing the expected log-likelihood, Eq.~\eqref{eq:EloglikeFinal} to find the next value of the parameter is now trivial: 
\begin{align}\label{eq:EMupdate}
    \vb*{\Sigma}[k+1]=\frac{1}{n_s+n_r}\qty(n_s\widehat{\vb*{\Sigma}} + n_r\vb*{A}[k])
\end{align}

\subsubsection{Inclusion of an Inverse-Wishart prior for $\vb*{\Sigma}$}
The generalization to maximizing the posterior for $\vb*{\Sigma}$ with a conjugate prior taking the Inverse-Wishart form is immediate. Taking $\vb*{\Psi}$ to be the parameter of the prior, $P=2\dim(s)$, and $\nu>P-1$ the number of degrees of freedom, then this modifies the EM update, Eq.~\eqref{eq:EMupdate} to 
\begin{align}
    \vb*{\Sigma}[k+1]=\frac{1}{n_s+n_r+(\nu+P+1)}\qty(n_s\widehat{\vb*{\Sigma}} + n_r\vb*{A}[k] + \vb*{\Psi})
    \label{eq:EMupdateMAP}
\end{align}
When $n_s\approx P$, the Maximum A Posteriori (MAP) estimator is quite different from the ML estimator. 

\subsubsection{Proof that EM iterations conserve positive (semi-) definiteness of $\vb*{\Sigma}$}
To prove the positive definiteness of the estimated covariance matrix, we recall the following very useful  characterization of \textit{positive semi-definite} (\textit{psd})  matrices using the Schur complement:\\
\textbf{Lemma} (e.g., \citet{Gallier_2011}):
Let $\vb*{M}_{22}$ be positive definite, $\vb*{M}_{22}>0$. Then 
\begin{align}
    \vb*{M}=\mqty(\vb*{M}_{11} & \vb*{M}_{12}\\ \vb*{M}_{12}^T & \vb*{M}_{22})\geq 0
\end{align}
if and only if $(\vb*{M} / \vb*{M}_{22})\geq 0$.

We wish to show that as long as we have enough surrogates such that covariance matrix estimated from them is positive definite, then it is true that if we initialize $\vb*{\Sigma}[0]$ such that $(\vb*{\Sigma}[0]/\vb*{\Sigma_{rr}}[0])\geq 0$ then $\vb*{\Sigma}[k] \geq 0$ throughout the EM iteration and therefore also for the fixed point.
This follows directly from Lemma 1, as follows.

At step $k$ of the EM iteration assume $\vb*{\Sigma}[k]$ is such that $(\vb*{\Sigma}[k]/\vb*{\Sigma_{rr}}[k])\geq 0$.
By assumption we always have enough surrogates that $\vb*{A}_{rr}>0$. Therefore $\vb*{A}_{rr}$ is invertible and we have that 
\begin{align}
    (\vb*{A}/\vb*{A}_{22})=(\vb*{\Sigma}[k]/\vb*{\Sigma_{rr}}[k])\geq \vb*{A}
\end{align}
by assumption. This implies $\vb*{A}>0$ by the Lemma.  The sum of two \textit{psd} matrices is itself \textit{psd}, and since $\widehat{\vb*{\Sigma}}$ is manifestly \textit{psd}, this guarantees that $\vb*{\Sigma}[k+1]\geq 0$. The "only if" direction of the Lemma guarantees that $(\vb*{\Sigma}[k+1]/\vb*{\Sigma_{rr}}[k+1])\geq 0$ at the next iteration. Therefore, $\vb*{\Sigma}[k]\geq 0$ for all $i\geq0$ by induction. 

\subsubsection{The Maximum Likelihood and A Posteriori solutions as Fixed Point of the EM iterations}

While the iterations are computationally very light, since we have closed-form solutions for the iterative updates (sections \ref{sec:ML} and \ref{sec:MAP}), we can do even better by deriving a closed-form solution directly for the iterative fixed point and thus demonstrate the equivalence with EM. Solving $\vb*{\Sigma}[k+1] = \vb*{\Sigma}[k] \equiv \widehat{\vb*{\Sigma}}^{\text{MAP}}$ by combining equation \eqref{eq:EMupdate} with equations \eqref{eq:Ass}, \eqref{eq:Asr} and \eqref{eq:Arr} gives

\begin{align}
    &\vb*{\widehat{\Sigma}_{rr}}^\text{EM} = \frac{n_r\vb*{A_{rr}} + (n_s + n_p)\vb*{\widehat{\Sigma}_{rr}^{\Delta}}}{n_s + n_r + n_p}\\
    &\widehat{\vb*{\Sigma_{sr}}}^\text{EM} = (n_s + n_p)\vb*{\widehat{\Sigma}_{sr}^{\Delta}}\\
    &\times \left( (n_s + n_r + n_p)\unit_{p_s} - [\widehat{\vb*{\Sigma}_{rr}}^\text{EM}]^{-1}n_r\vb*{A_{rr}} \right)\\
    &\vb*{\widehat{B}^{\text{EM}}} = \widehat{\vb*{\Sigma_{sr}}}^\text{EM} [\widehat{\vb*{\Sigma}_{rr}}^\text{EM}]^{-1}\\
    &\widehat{\vb*{\Sigma_{ss}}}^\text{EM} = \frac{(n_s + n_p)\vb*{\widehat{\Sigma}_{sr}^{\Delta}} + n_r  \vb*{\widehat{B}^{\text{EM}}}  (\vb*{A_{rr}} - \vb*{\widehat{\Sigma}_{rr}}^\text{EM})\vb*{\widehat{B}^{\text{EM}}}^T}{n_s + n_p}\label{eq:CovSSEM}
\end{align}

Equation \eqref{eq:CovSSEM} is equivalent to equation \eqref{eq:covSSMAP}, even though it looks more complicated. This is because solving for the EM introduces the covariance of the unpaired surrogates only $\vb*{A_{rr}}$ and not the covariance of the paired and unpaired surrogates $\vb*{\widehat{\Sigma}_{rr}^{\star}}$.

\subsection{Case when the surrogate covariance is known}
\label{sec:knownSurrCov}
We can rewrite the simulation summary statistics covariance from section \ref{sec:regML} as
\begin{align}
    \widehat{\vb*{\Sigma}^\text{MAP}_{ss}} &= \vb*{\widehat{\Sigma}^{\Delta}_{ss}} + \frac{n_r}{n_r+n_s+n_p}
        \widehat{\vb{B}}_\text{MAP}
        \qty(\widehat{\vb*{\Sigma}_{rr}}-\widehat{\vb*{\Sigma}_{rr}}^{\Delta})
        \widehat{\vb{B}}_\text{MAP}^T\,. \label{eq:fullEstimator}
\end{align}
% \bdw{This explains why the estimator with $n_p=0$ goes crazy when the number of paired samples is small: $B$ contains the inverse of the ML estimate of the surrogate covariance from the paired samples only. This didn't cause the code to break, because the formulas hid this and we never actually divided by a formally degenerate matrix.}
which  is strictly equivalent to equation \eqref{eq:covSSMAP}. 
%This form looks remarkably like an estimate based on $n_s+n_p$ terms from the first summand and $n_r$ terms from the second summand. It combines the $n_r$ surrogates in $\widehat{\vb*{\Sigma}}_{rr}$ with the $n_s+n_p$ samples in $\widehat{\vb*{\Sigma}}_{rr}^{\Delta}$ and then rescales the sum  as if there were at total of $n_r$ terms in the second summand.

The case when a theoretical covariance $\vb*{\Sigma}_{rr}$ for the surrogates is available directly obtains from  the limit of Eq.~\eqref{eq:fullEstimator} as $n_r\rightarrow \infty$ 
\begin{align}
       \vb*{\widehat{\Sigma}_{ss}^{\scaleto{\mathrm{MAP},\vb*{\Sigma}_{rr}}{5.0pt}}} &=
        \lim_{n_r\rightarrow \infty}\vb*{\widehat{\Sigma}_{ss}^{\scaleto{\mathrm{MAP}}{3.5pt}}} =\nonumber\\
        &\vb*{\widehat{\Sigma}^{\Delta}_{ss}} + 
        \widehat{\vb{B}}_\text{MAP}
        \qty(\vb*{\Sigma}_{rr}-\widehat{\vb*{\Sigma}}_{rr}^{\Delta})
        \widehat{\vb{B}}_\text{MAP}^T.
\end{align}

\section{MAP derivation (regression parameters)}\label{app:regressionMAP}

This section presents the derivation of the closed-form solutions for the covariance in section \ref{sec:newMAP}.
We can extend Anderson's in \citet{Anderson1957} derivation by including an Inverse-Wishart Prior with parameters $\vb*{\Psi}$ and $\nu$. Under the hypothesis that the block covariance $\Sigma$ of simulation and surrogates summary statistics is drawn from and Inverse-Wishart distribution (equation \eqref{eq:invWishart}, the following properties hold true: 
 \begin{enumerate}
     \item $\vb*{\Sigma_{rr}} \indep \vb*{\Sigma_{rr}}^{-1}\vb*{\Sigma_{rs}} = \vb*{B^T}$.
      \item $\vb*{\Sigma_{rr}} \indep \vb*{\Sigma_{s|r}}$.
      \item $\vb*{\Sigma_{rr}} \sim \mathcal{W}^{-1}(\vb*{\Psi_{rr}}, \nu - p_s)$.
      \item $\vb*{\Sigma_{s|r}} \sim \mathcal{W}^{-1}(\vb*{\Psi_{s|r}}, \nu)$ with $\vb*{\Psi_{s|r}} \equiv (\vb*{\Psi}/\vb*{\Psi_{rr}})$.
      \item $\vb*{B^T}|\vb*{\Sigma_{s|r}} \sim \mathcal{MN}(\vb*{\Psi_{rr}}^{-1}\vb*{\Psi_{rs}}, \vb*{\Sigma_{s|r}}\otimes \vb*{\Psi_{rr}}^{-1})$ where $\otimes$ is the Kronecker product and $\mathcal{MN}$ designates the matrix normal distribution.
 \end{enumerate}
This is particularly convenient for our problem and we can extend Anderson's result straightforwardly to a Maximum A Posteriori (MAP) estimate. In particular, we can re-parametrize the distribution
\begin{equation}
    \mathcal{P}(\vb*{\Sigma}) = \mathcal{P}(\vb*{B^T}|\vb*{\Sigma_{s|r}} )\mathcal{P}(\vb*{\Sigma_{s|r}})\mathcal{P}(\vb*{\Sigma_{rr}})\,.
\end{equation}

Let's index the unpaired surrogate samples as $\vb*{r}_i$ with $i=1,\dots, n_r$ and the surrogate samples that are part of the pairs $\vb*{x}_i$ $i=1,\dots,n_s$ as $\vb*{r}_i$ with $i=n_r+1,\dots,n_r + n_s$.
We factorize the likelihood as Anderson, that is to say
\begin{align}
    \mathcal{L}(\{\vb*{x}\},\{\rstar\}| \vb*{\Sigma}) &=  \prod_{i=1}^{n_s} \mathcal{P}(\vb*{x}_i|\vb*{\mu_s},\vb*{\Sigma_{ss}}) \prod_{j=1}^{n_r} \mathcal{P}(\vb*{r}_j|\vb*{\mu_r},\vb*{\Sigma_{rr}})\\
    &= \prod_{i=1}^{n_s+n_r}\mathcal{P}(\vb*{r}_i|\vb*{\mu_r}, \vb*{\Sigma_{rr}})\prod_{i=1}^{n_s}\mathcal{P}(\vb*{s}_i|\vb*{\mu}_{\vb*{s}_i|\vb*{r}_i},\vb*{\Sigma_{s|r}})\nonumber
\end{align}
The right hand side depends separately on $\vb*{\Sigma_{rr}}$, $\vb*{\Sigma_{s|r}}$ and $\vb*{B^T}$ (through $\vb*{\mu_{s|r}}$) as the prior, so we can solve the MAP problem from equation \eqref{eq:MAPproblem}.

Then the natural logarithm posterior distribution is

\begin{align}
     &-2\lnb{\mathcal{P}(\vb*{\Sigma}|\{\vb*{x}\},\{\rstar\})}= (n_s + n_r + \nu -p_s + p_r + 1) \lnb{\mathrm{det}(\vb*{\Sigma_{rr}})}\nonumber\\
     &+ (n_s + \nu + 2p_s + 1)\lnb{\mathrm{det}(\vb*{\Sigma_{s|r}})}\nonumber \\
     &+ \mathrm{Tr} \Bigg( \bigg[ \sum_{i=1}^{n_s+n_r}(\vb*{r}_i - \vb*{\mu_r})(\vb*{r}_i - \vb*{\mu_r})^{\vb*{T}} + \vb*{\Psi_{rr}} \bigg] \vb*{\Sigma_{rr}}^{-1}\Bigg) \nonumber \\ &+ \mathrm{Tr} \Bigg( \bigg[ \sum_{j=1}^{n_s}(\vb*{s}_j - \vb*{\mu_{s|r}})(\vb*{s}_j - \vb*{\mu_{s|r}})^{\vb*{T}} + \vb*{\Psi_{s|r}} \nonumber\\
     &+  (\vb*{B^T} - \vb*{\Gamma^T})^{\vb*{T}}\vb*{\Psi_{rr}}(\vb*{B^T} - \vb*{\Gamma^T})\bigg] \vb*{\Sigma_{s|r}}^{-1} \Bigg)
\end{align}

We then solve successively $\frac{\partial\lnb{\mathcal{P}(\vb*{\Sigma}|\{\vb*{x}\},\{\rstar\})}}{\partial \alpha} = 0$ for $\alpha \in \left\{\vb*{\Sigma_{rr}}, \vb*{B^T}, \vb*{\Sigma{s|r}}  \right\}$, after a bit of derivation and linear algebra, and we find the solutions from section \ref{sec:newMAP}, i.e. equations \eqref{eq:regRR}, \eqref{eq:regB} and \eqref{eq:regSR} which allow to compute $\vb*{\widehat{\Sigma}_{ss}^{\scaleto{\mathrm{MAP}}{3.5pt}}}$ from equation \eqref{eq:covSSnewMAP}

\section{Some additional results}\label{app:moreResults}

\subsection{Relative gain for the power spectrum}\label{app:morePk}
\begin{figure*}
    \includegraphics[width=\textwidth]{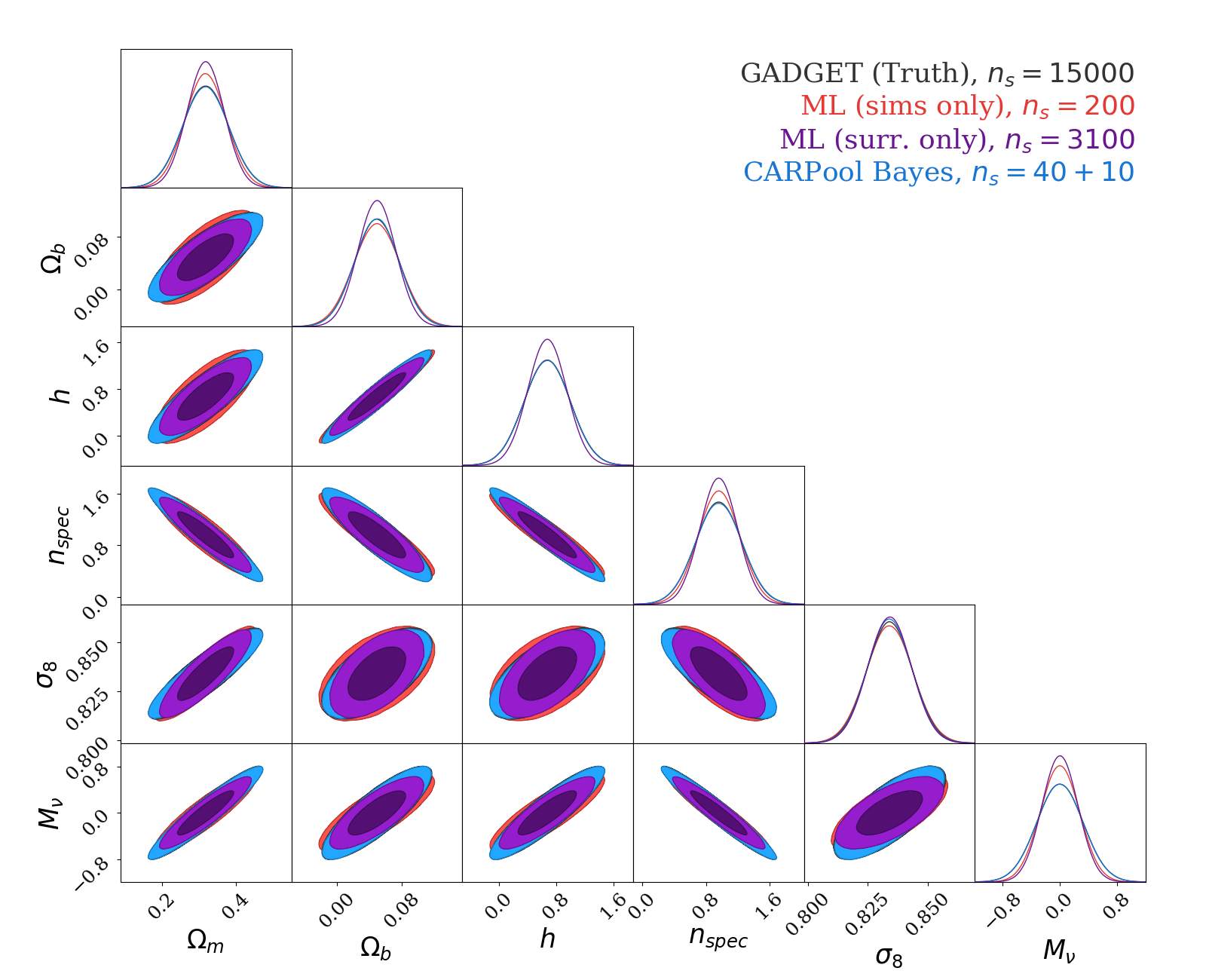}
    \caption{Fisher confidence contours of the cosmological parameters based on the estimated  covariance matrix of the matter power spectrum. The estimators which we compare are the same as in Figure \ref{fig:fisherPk}, except that we have now $n_s = 40 + 10$ simulations for CARPool Bayes (empirical Bayes prior on the block covariance).}
    \label{fig:fisherPk2}
\end{figure*}

We simply show the confidence bounds for the $\Lambda$CDM parameters using the power spectrum covariance matrix, this time with more simulations for the CARPool Bayes covariance, i.e $n_s = 40 + 10$ versus $n_s = 10 + 5$ in section \ref{sec:illustration}. Figure \ref{fig:fisherPk2} shows CARPool Bayes marginal bounds even closer to the truth than in Figure \ref{fig:fisherPk} at the price of running $50$ simulations in total instead of $15$. This demonstrates the relative gain of running more simulations is small for the covariance matrix when the simulation and surrogate summary statistics are well correlated. 

\subsection{MAP on the regression parameters}\label{sec:mapRegExp}
\begin{figure*}
    \includegraphics[width=\textwidth]{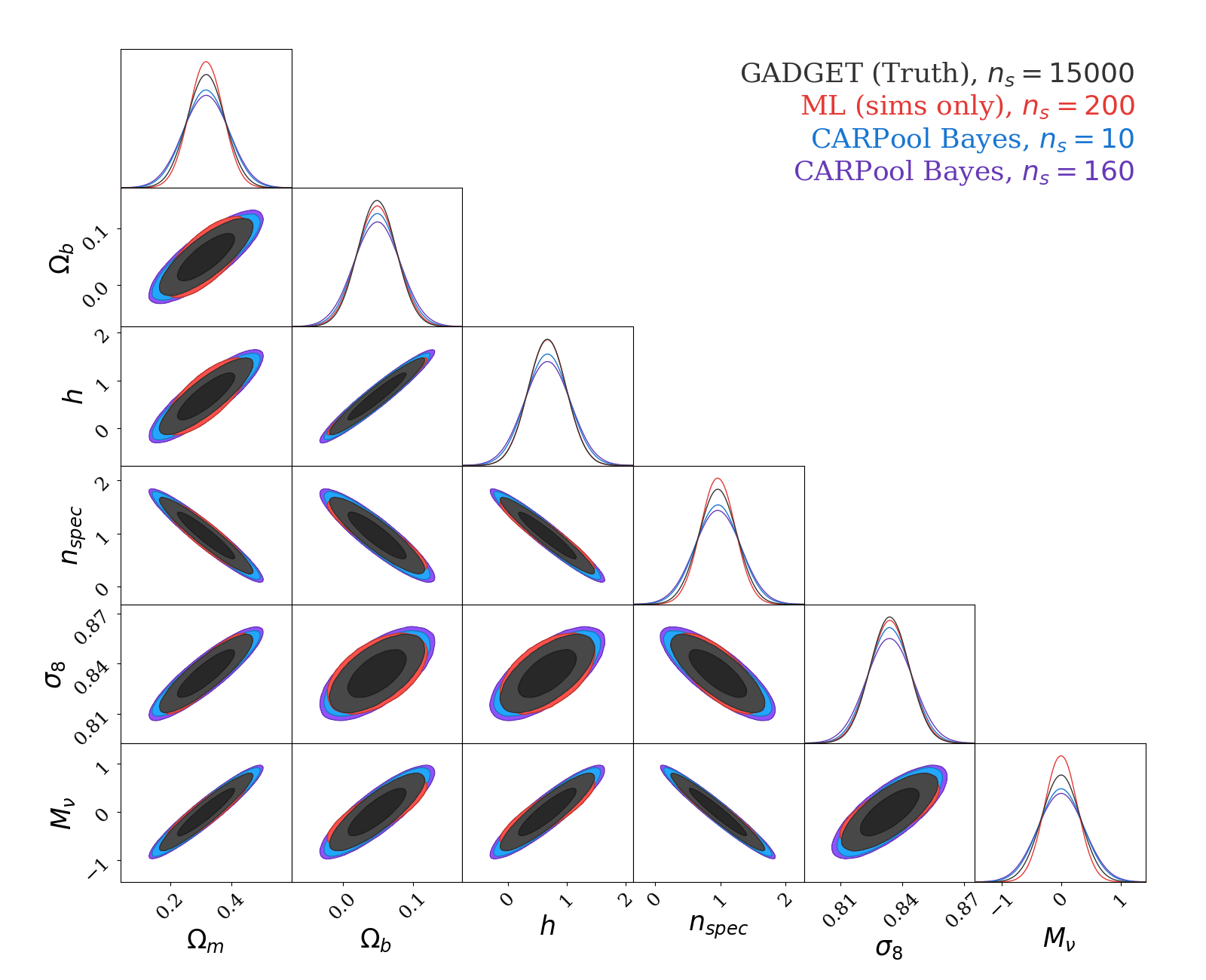}
    \caption{Confidence contours of the cosmological parameters computed using the Fisher matrix based on the estimated matter power spectrum covariance The "CARPool Bayes" estimates follow the computations of section \ref{sec:newMAP}, where the prior, still the empirical Bayes one from section \ref{sec:empBayes}, is parametrized given the regression parameters. We stress that this is the only Figure in the paper that shows a computation of the "regression" MAP from section \ref{sec:newMAP}.}
    \label{fig:fisherPkMyMAP}
\end{figure*}

We chose to present one particular example of the MAP estimate from section \ref{sec:newMAP} on the power spectrum, which showed the most successful results with the "block" parametrization from section \ref{sec:regML} with only $n_s = 10 + 5$ simulations.
We fix $\nu = 2*p_s + 2$ in this case and do not consider it a free parameter, nor do we allow it to define an improper prior, i.e. we do not allow $\nu \leq 2*ps -1$. This corresponds to the lowest integer for which the expectation of the Inverse-Wishart exists.
In Figure \ref{fig:fisherPkMyMAP}, the marginal confidence bounds are much wider than the truth for both $n_s = 10$ and $n_s = 160$ for the CARPool Bayes estimator (this time the "regression" framework from section \ref{sec:newMAP}). Since the MAP on the regression parameters does not allow for an improper prior, the estimator of the simulation covariance puts too much weight on the na\"ive diagonal empirical Bayes prior we use (section \ref{sec:empBayes}). For future studies, we can explore whether having a "smarter prior", for instance a model covariance computed from theoretical approximations to parametrize the Inverse-Wishart distribution, can significantly improve or not both the CARPool Bayes estimators from sections \ref{sec:regML} and \ref{sec:newMAP}.

\subsection{Results from the bispectrum}\label{app:Bk}
Here, we directly present the confidence bounds for the $\Lambda$CDM parameter found using various covariance estimators of the bispectrum.
The motivation here is to demonstrate the improvement over \carpcov\ for the same summary statistics.
The first summary statistics we test is the set of squeezed isosceles triangles, that is to say the bispectra computed for $k_1=k_2$ and in ascending order of the ratio $k_3/k_1 \leq 0.20$ ($p_s = 98$ in this case).

\begin{figure*}
    \includegraphics[width=\textwidth]{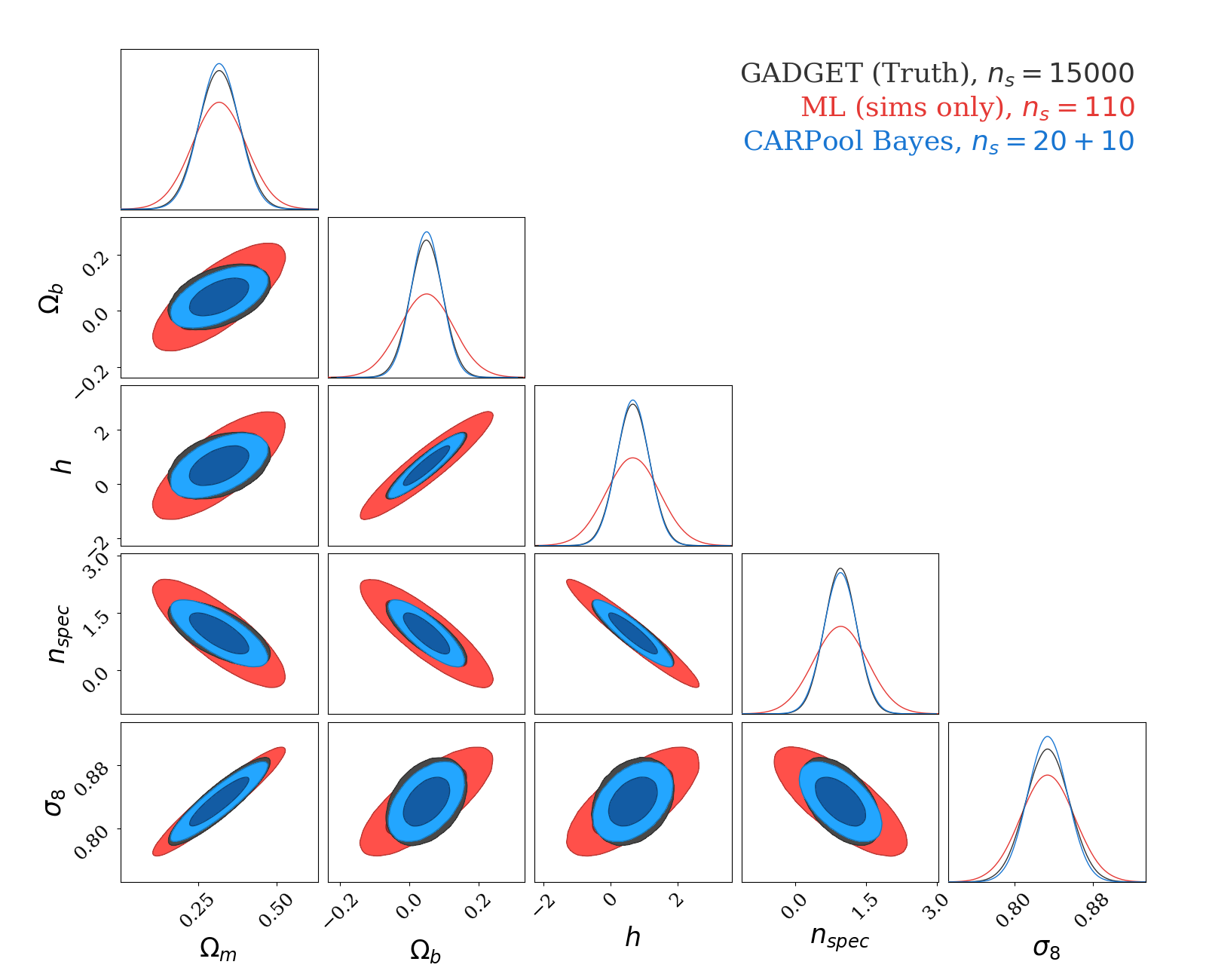}
    \caption{Confidence contours of the cosmological parameters computed using the Fisher matrix based on the estimated matter bispectrum covariance matrix, for a set of squeezed isosceles triangles. The estimators result from the same computations as in Figure \ref{fig:fisherPk}}.
    \label{fig:fisherBkSq}
\end{figure*}

Figure \ref{fig:fisherBkSq} demonstrate that we get parameter constraints much more representative of the truth with $n_s = 20 + 10$ simulations that with the sample covariance using $n_s = 110$ simulations. The CARpool Bayes estimator is the one from section \ref{sec:regML} using the empirical Bayes prior from section \ref{sec:empBayes}.

\begin{figure*}
    \includegraphics[width=\textwidth]{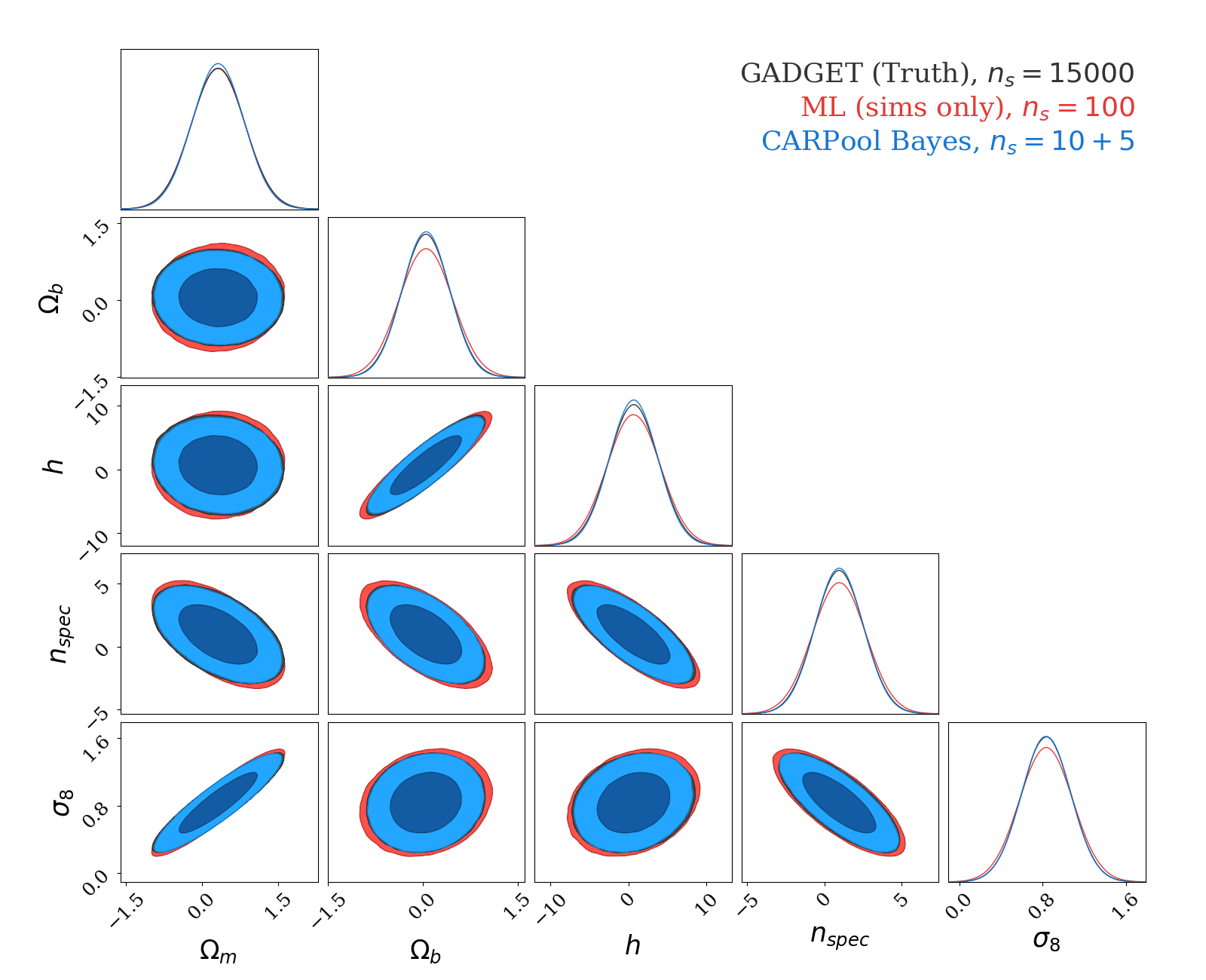}
    \caption{Confidence contours of the cosmological parameters computed using the Fisher matrix based on the estimated matter bispectrum covariance matrix, for a set of equilateral triangles. The estimators we compare are the same as in Figure \ref{fig:fisherPk}}
    \label{fig:fisherQkEq}
\end{figure*}

Then, we take a look at the reduced bispectrum of equilateral equilateral triangles with $k_1=k_2=k_3$ varying up to $k_\mathrm{max} = 0.75$ $h{\rm Mpc^{-1}}$ ($p_s=40$). In Figure \ref{fig:fisherQkEq}, we observe the CARPool Bayes estimator gives almost identical parameters marginal contours to the truth with only $n_s = 10 + 5$ simulations, while the sample covariance of simulations uses $n_s = 100$ simulations.

\bibliographystyle{mnras}
\bibliography{covBib}

% Don't change these lines
\bsp	% typesetting comment
\label{lastpage}
\end{document}